\documentclass[prl,superscriptaddress,twocolumn]{revtex4-1}
\usepackage{tikz}
\usepackage{graphicx}
\usepackage{amsmath}
\usepackage{gensymb}
\usepackage[margin=1cm]{geometry}
\usepackage{setspace}
\usepackage{xcolor}
\usepackage{cancel}
\usepackage{soul}
\usepackage{wrapfig}
\usepackage{ulem}
\usepackage{units}
\usepackage{lipsum}

\usepackage[colorlinks,urlcolor=blue,citecolor=blue,linkcolor=blue]{hyperref}

\newcommand{\reftextit}[1]{}

\begin{document}

\title{Valley-mediated singlet- and triplet-polaron interactions and quantum dynamics\\ in a doped WSe$_2$ monolayer}

\author{Yue Ni}
\thanks{These authors contributed equally to this work.}
\affiliation{Department of Physics and Center for Complex Quantum Systems, 
    The University of Texas at Austin, Austin, TX, 78712, USA}
\author{Di Huang}
\thanks{These authors contributed equally to this work.}
\affiliation{Department of Physics and Center for Complex Quantum Systems, 
    The University of Texas at Austin, Austin, TX, 78712, USA}
\author{Danfu Liang}
\affiliation{Department of Physics and Center for Complex Quantum Systems, 
    The University of Texas at Austin, Austin, TX, 78712, USA}
\author{Albert Liu}
\affiliation{Condensed Matter Physics and Materials Science Division, Brookhaven National Laboratory, Upton, New York, 11973, USA}
\author{Xiaohui Liu}
\affiliation{Department of Physics and Center for Complex Quantum Systems, 
    The University of Texas at Austin, Austin, TX, 78712, USA}
\author{Kevin Sampson}
\affiliation{Department of Physics and Center for Complex Quantum Systems, 
    The University of Texas at Austin, Austin, TX, 78712, USA}
\author{Zhida Liu}
\affiliation{Department of Physics and Center for Complex Quantum Systems, 
    The University of Texas at Austin, Austin, TX, 78712, USA}
\author{Jianmin Quan}
\affiliation{Department of Physics and Center for Complex Quantum Systems, 
    The University of Texas at Austin, Austin, TX, 78712, USA}
\author{Kenji Watanabe}
\affiliation{Research Center for Electronic and Optical Materials, National Institute for Materials Science, 1-1 Namiki, Tsukuba 305-0044, Japan}
\author{Takashi Taniguchi}
\affiliation{Research Center for Materials Nanoarchitectonics, National Institute for Materials Science,  1-1 Namiki, Tsukuba 305-0044, Japan}
\author{Dmitry K. Efimkin}
\affiliation{School of Physics and Astronomy and ARC Centre of Excellence in Future Low-Energy Electronics Technologies, Monash University, Victoria 3800, Australia}
\author{Jesper Levinsen}
\affiliation{School of Physics and Astronomy and ARC Centre of Excellence in Future Low-Energy Electronics Technologies, Monash University, Victoria 3800, Australia}
\author{Meera M. Parish}
\affiliation{School of Physics and Astronomy and ARC Centre of Excellence in Future Low-Energy Electronics Technologies, Monash University, Victoria 3800, Australia}
\author{Xiaoqin Li}%
\email{elaineli@physics.utexas.edu}
\affiliation{Department of Physics and Center for Complex Quantum Systems, 
    The University of Texas at Austin, Austin, TX, 78712, USA}
\date{\today}

\begin{abstract}

In doped transition metal dichalcogenides, optically created excitons (bound electron-hole pairs) can strongly interact with a Fermi sea of electrons to form Fermi polaron quasiparticles. When there are two distinct Fermi seas, as is the case in WSe$_2$, there are two flavors of lowest-energy (attractive) polarons---singlet and triplet---where the exciton is coupled to the Fermi sea in the same or opposite valley, respectively. Using two-dimensional coherent electronic spectroscopy, we analyze how their quantum decoherence evolves with doping density and determine the condition under which stable Fermi polarons form.  
Because of the large oscillator strength associated with these resonances, intrinsic quantum dynamics of polarons as well as valley coherence between coupled singlet- and triplet polarons occur on sub-picosecond time scales. Surprisingly, we find that a dark-to-bright state conversion process leads to a particularly long-lived singlet polaron valley polarization, persisting up to \unit[200-800]{ps}. Valley coherence between the singlet- and triplet polaron is correlated with their energy fluctuations. Our finding provides valuable guidance for the electrical and optical control of spin and valley indexes in atomically thin semiconductors. 
\end{abstract}


\maketitle

The concept of the polaron, where a mobile impurity is modified by a surrounding quantum medium, is fundamental in physics. First introduced in the solid-state context to describe how electrons become dressed by phonon excitations in a crystal lattice, it has since found applications from the densest nuclear matter to the most dilute cold atomic gases~\cite{2014_Massignan_RepProgPhys_Review, FermiGasesReview}. More recently, Fermi polarons have been proposed to describe new quasiparticles when optically created excitons are coherently coupled to a Fermi sea of electrons or holes in doped van der Waals semiconductors, in particular, transition metal dichalcogenides (TMDs)~\cite{2016_A.Imamoglu_NatPhys_MoSe2, 2017_Dmitry_PRB_theory,2019_Dmitry_PRB_theory, muir_interactions_2022}. The attractive and repulsive polarons in TMDs evolve into neutral excitons and trions (charged exciton bound states), respectively, in the limit of vanishing doping~\cite{2016_A.Imamoglu_NatPhys_MoSe2,2017_Dmitry_PRB_theory,huang_quantum_2023}.

\begin{figure}
    \centering
    \includegraphics[width=8.6cm]{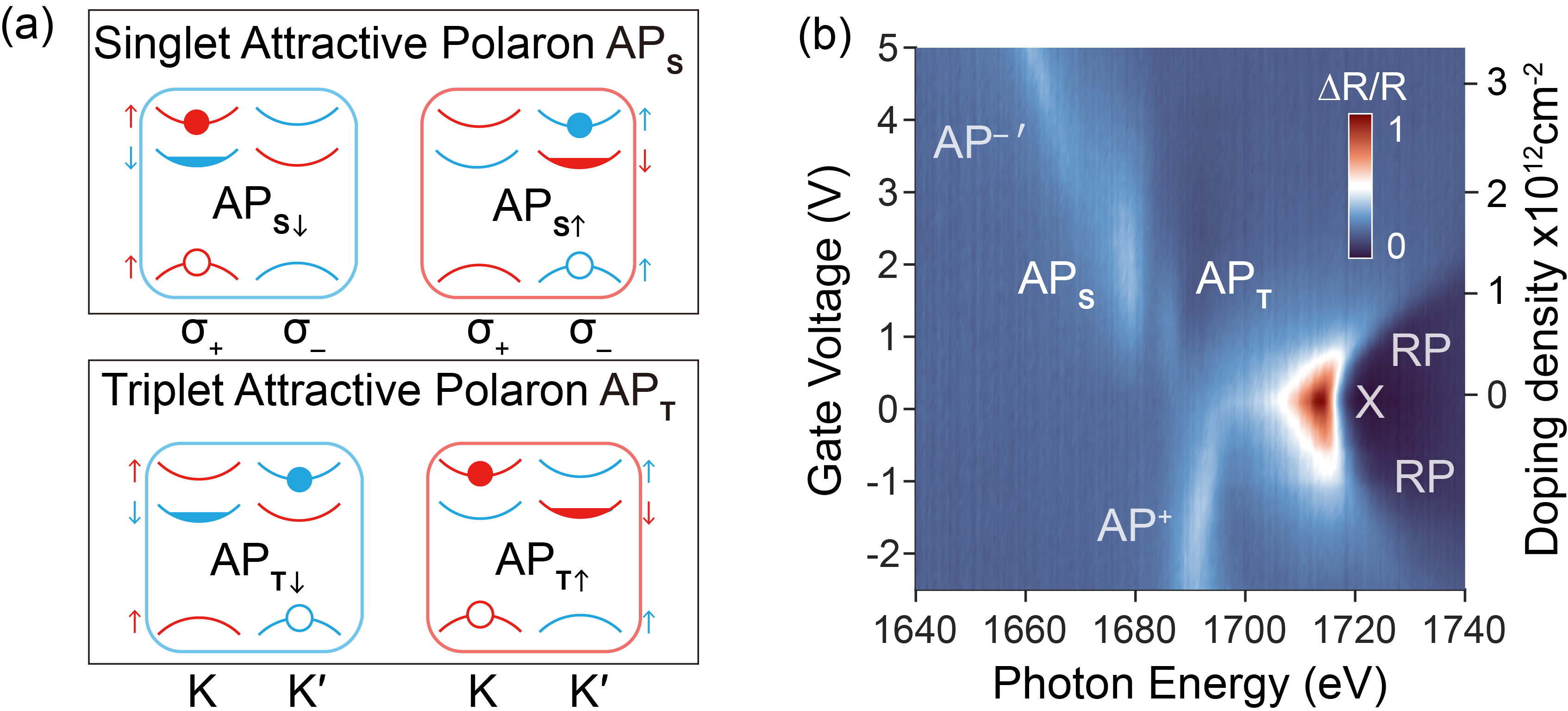}
     \caption{Polaron resonances in a doped WSe$_2$ monolayer.
     (a) Schematic of the singlet and triplet attractive polarons in a WSe$_2$ monolayer. Here,
     AP$_{S\downarrow}$ (AP$_{S\uparrow}$) refers to an $\sigma_{+}$($\sigma_{-}$) circularly polarized light excited exciton in the $K$ ($K'$) valley coupled to a Fermi sea in the same valley with spin down (up) denoted by blue (red). For triplet polarons, excitons are coupled to doped electrons in the opposite valley. (b) Reflectance spectrum as a function of gate voltage (left axis), i.e., doping density (right axis). Different attractive and repulsive polarons (RP) are identified. AP$^+$: hole-doped AP; AP$_S$/AP$_T$: electron-doped singlet/triplet attractive polarons; AP$^-{}'$: a higher-order many-body AP. 
    }
    \label{fig1}
\end{figure}

Polarons in TMDs exhibit complex quantum dynamics and interaction effects as the electrons and excitons acquire an additional valley index. Due to spin-valley locking, one can selectively create excitons in one of the two valleys ($K$ and $K'$) at the corners of the hexagonal Brillouin zone by choosing the helicity of the circularly polarized excitation light~\cite{2018_G.Wang_RMP_ExcitonReview,xiao_coupled_2012}. 
In  WX$_2$ (X=S, Se) monolayers, the lowest-energy bright exciton consists of an electron in the second lowest conduction band and a hole in the highest valence band. 
When electron dopants are introduced in both WS$_2$ and WSe$_2$, two types of attractive polarons, singlet- and triplet- attractive polarons, form when the optically created excitons are dressed by the Fermi sea in the same (opposite) valley as illustrated in Fig.~\ref{fig1}a.

How quantum dynamics of polarons evolve with doping density remain important yet open questions because it serve as a critical criterion to distinguish different theoretical frameworks for describing fundamental optical excitations in doped TMDs~\cite{Vaclavkova_2018}. Here, we address these questions by studying a doped WSe$_2$ monolayer using two-dimensional coherent electronic spectroscopy (2DCES) with a variety of pulse sequences and polarization schemes.  
We find that, with increasing doping density, the quantum dephasing rate of singlet and triplet polarons initially remains constant, revealing the condition under which stable Fermi polarons form. With higher doping density, the triplet polaron dephasing increases faster than that of singlet polarons. Valley decoherence between coupled singlet- and triplet polarons is found to correlate with their energy fluctuations because of the shared Fermi sea.
Surprisingly, we observe a long-lived valley polarization associated with both types of attractive polarons, where the singlet polarons exhibit the most robust valley polarization.
We attribute this slow relaxation to the band inversion in WSe$_2$ monolayers, which enables a dark-to-bright state conversion process~\cite{jadczak_upconversion_2021,danovich_dark_2017,cong_interplay_2023} that is absent in doped MoSe$_2$ monolayers.

We control the doping density in a pristine WSe$_2$ monolayer embedded in a device consisting of hBN encapsulation layers, a few-layer graphite top gate (TG), and a metal back gate (details in the Supplementary Information (SI)~\cite{SI}). All optical measurements are performed at \unit[10]{K}.
\nocite{2013_G.Nardin_OptExp_2DCSSetup, 2015_G.Moody_NatComm_2DCSMoSe2, 2015_Kormanyos_2DMat_MoSe2Review, D.Hanan_DBTheory_2020, robert_measurement_2020}
The reflectance spectra as a function of top gate voltages ($V_{TG}$) feature several resonances as shown in Fig.~\ref{fig1}b. In the intrinsic region with minimal doping~(i.e., $V_{TG} < \unit[0.2]{V}$), the spectrum is dominated by a sharp neutral exciton (X) resonance. When doping density increases, excitons evolve into repulsive polarons that exhibit a continuous blue shift of the resonant energy~\cite{2020_Glazov_JCP_EqualTrionPolaron,TrionDoubtSuris,TrionDoubtWouters2}. 
At the same time, additional peaks corresponding to attractive polarons appear at lower energies than the neutral exciton and are continuously connected to bound trions, or charged excitons, at vanishing doping density. 

While there is only one attractive polaron branch (AP$^+$) with hole doping, two attractive polarons, singlet- and triplet-attractive polarons (AP$_S$ and AP$_T$) are observed with electron doping~\cite{2020_A.Chernikov_PRL_WSe2Trion,2021_CHLui_NatComm_MoSe2&WSe2}. The simple band picture in Fig.~\ref{fig1}a does not include exchange interactions between bands~\cite{yu_dirac_2014,lyons_valley_2019} and thus does not capture the observation that the singlet state is lower in energy than the triplet state by $\sim$\unit[7]{meV}. At very high electron doping density, another resonance (AP$^{-'}$) emerges and rapidly shifts to lower energy with increasing electron doping density. 
This resonance has been attributed to a higher-order many-body state in a previous study~\cite{PhysRevLett.129.076801}.

The AP$_S$ (AP$_T$) resonance consists of an exciton coupled to a Fermi sea in the same (opposite) valley. 
We introduce one more subscript ($\uparrow$ and $\downarrow$) to label the electron spin of the Fermi sea. 
There are thus four different attractive polaron states (AP$_{S \downarrow}$, AP$_{T\uparrow}$, AP$_{S\uparrow}$, AP$_{T\downarrow}$) for singlet (intra-valley) and triplet (inter-valley) coupling to electrons in the $K$ and $K$' valleys, as shown in Fig.~\ref{fig1}a. While we do not label the valley index explicitly to simplify the notation, the transitions and notations are summarized in Fig.~\ref{fig1}a, in which $\sigma_{+}$($\sigma_{-}$) circularly polarized light creates excitons in the $K$ ($K'$) valley.

\begin{figure}
    \centering
    \includegraphics[width=8.6cm]{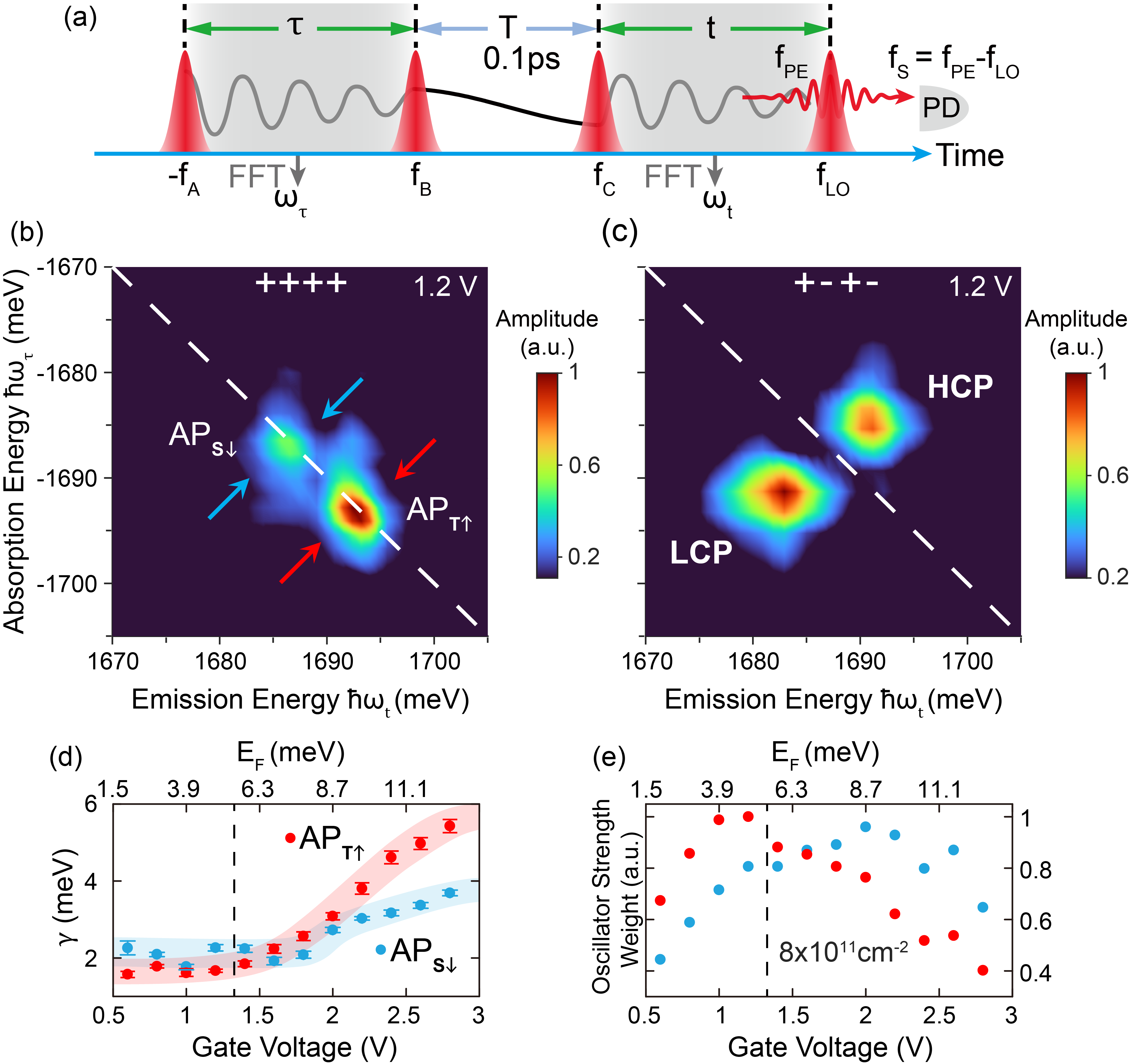}
    \caption{Co-circular one-quantum rephasing spectra.
    (a) The one-quantum rephasing pulse sequence. The time domain signals as a function of delays $\tau$ and $t$ are Fourier transformed to obtain the absorption and emission frequency axes ($\hbar\omega_\tau$ and $\hbar\omega_t$).
    (b) Normalized 2D amplitude spectrum taken at \unit[1.2]{V} top gate voltage using ++++ polarization.
    The two diagonal peaks correspond to singlet and triplet attractive polarons in the $K$ valley (AP$_{S\downarrow}$ and AP$_{T\uparrow}$).
    (c) Normalized 2D amplitude spectrum obtained using cross-circular polarization (+$-$+$-$) at \unit[1.2]{V}. Only two off-diagonal higher (HCP) and lower (LCP) cross peaks are observed, corresponding to the interaction between AP$_{S}$ and AP$_{T}$.
    (d) Homogeneous linewidth $\gamma$ and (e) oscillator strength weight of AP$_{S\downarrow}$ and AP$_{T\uparrow}$ as a function of gate voltage (e.g., doping density) and corresponding Fermi energy. The black dashed line indicates the doping density $n_{e^-}=8\times10^{11}$ cm$^{-2}$.
    }
    \label{fig2}
\end{figure}

To study quantum dynamics and interactions of AP$_S$ and AP$_T$, we use the collinear 2DCES setup~\cite{nardin_multi-dimensional_2015} depicted in the SI~\cite{SI}. 
Three $\sim$\unit[100]{fs} pulses at \unit[80]{MHz} repetition rate derived from a Ti:Sapphire laser are focused via a microscope objective lens to the same spot with a $\sim$ 1$\mu$m diameter. 
The emitted photon-echo signal is collected in the reflection geometry via the same objective lens and combined with a reference pulse for heterodyne detection. The power of each beam is kept at \unit[10]{$\mu$W} unless otherwise stated, corresponding to an exciton density of $n_{x}=8\times10^{10}\mathrm{cm}^{-2}$. Both the high-quality sample and the tightly focused spot in the collinear geometry are essential to spectrally resolve the singlet- and triplet states in the 2D spectra, which is critical to reveal their different dynamics and coupling.

We first perform one-quantum rephasing experiments where delays between the first two pulses ($\tau$) and between the third and the reference pulses ($t$) are scanned and then Fourier transformed to generate the absorption energy ($\hbar\omega_{\tau}$) and emission energy ($\hbar\omega_{t}$), as illustrated in Fig.~\ref{fig2}a. 
The waiting time $T$ between the second and third pulses is kept above \unit[0.1]{ps} to avoid coherent artifacts originating from nonlinear interaction terms during the temporal overlap of the pulses. The normalized amplitude spectrum with all co-circularly polarized pulses (++++) is shown in Fig.~\ref{fig2}b. At the doping density of $n_{e^-}=\unit[7\times10^{11}] {e^-/cm^2}$ and $V_\mathrm{TG}=\unit[1.2]{V}$, two resonances, AP$_{T\uparrow}$ and AP$_{S\downarrow}$, at \unit[1693]{meV} and \unit[1686]{meV} respectively, are observed along the diagonal (white dashed line). The absence of cross-peaks in Fig.~\ref{fig2}b suggests that AP$_{S\downarrow}$ and AP$_{T\uparrow}$ do not couple to each other because they do not share the same Fermi sea; thus, they are not competing for the same electrons. By contrast, the cross peaks dominate the one-quantum spectrum taken with cross-circular polarized pulses (+$-$+$-$ in Fig.~\ref{fig2}c) because singlet- and triplet polarons in opposite valleys share the same Fermi sea as illustrated in Fig.~\ref{fig1}a. These observations are consistent with those found in WS$_2$~\cite{muir_interactions_2022}, indicating that the polaron interaction mechanism is universal in TMDs.

We further perform a quantitative analysis of the quantum dynamics of AP$_{T\uparrow}$ and AP$_{S\downarrow}$ and their evolution as a function of doping density. 
The half width half maximum (HWHM) of a line profile along the cross diagonal direction indicated by arrows in Fig.~\ref{fig2}b reveals the intrinsic homogeneous linewidth $\gamma$~\cite{2010_Mark_OE_2DCSLineShape,2020_E.Martin_PRApp_2DCSMoSe2} or dephasing rate, which is inversely proportional to the quantum coherence time $T_2=\hbar/\gamma$. We extract $T_2=\unit[0.29]{ps}$ for AP$_S$ and $\unit[0.4]{ps}$ for AP$_T$ from Fig.~\ref{fig2}b. 
The dephasing rates $\gamma$ of both AP$_{T\uparrow}$ and AP$_{S\downarrow}$ remain constant with increasing electron doping density up to $\unit[8\times10^{11}] {e^-/cm^2}$ as shown in Fig.~\ref{fig2}d. Such robust quantum coherence supports our choice of adapting the polaron theory to describe these resonances~\cite{2021_Dmitry_PRB_theory}. At higher doping density, the dephasing rate of attractive polarons starts to increase due to the appearance of an additional decay channel to the higher-order many-body state AP$^-{}'$. We find that a Fermi polaron theory extended to two Fermi seas~\cite{SI}, while not capturing the AP$^-{}'$ state, correctly predicts the faster increase of the dephasing rate of AP$_{T}$ than AP$_{S}$ because of the decay from AP$_{T}$ to AP$_{S}$.

The relative oscillator strengths of AP$_{T\uparrow}$ and AP$_{S\downarrow}$ also evolve with doping density as shown in Fig.~\ref{fig2}e. Initially, at low density, both oscillator strengths increase linearly in accordance with both polaron and trion theories~\cite{Ngampruetikorn2012,2020_Glazov_JCP_EqualTrionPolaron}. At higher densities, the AP$_{T}$ oscillator strength reaches a maximum around $\unit[0.7\times10^{12}] {e^-/cm^2}$ before the oscillator strength is transferred to the energetically favorable AP$_{S}$. While this behavior cannot be captured within a trion theory, it is qualitatively reproduced by a polaron theory that accounts for the interactions with each of two Fermi seas~\cite{SI}.

\begin{figure}
    \includegraphics[width=8.6cm]{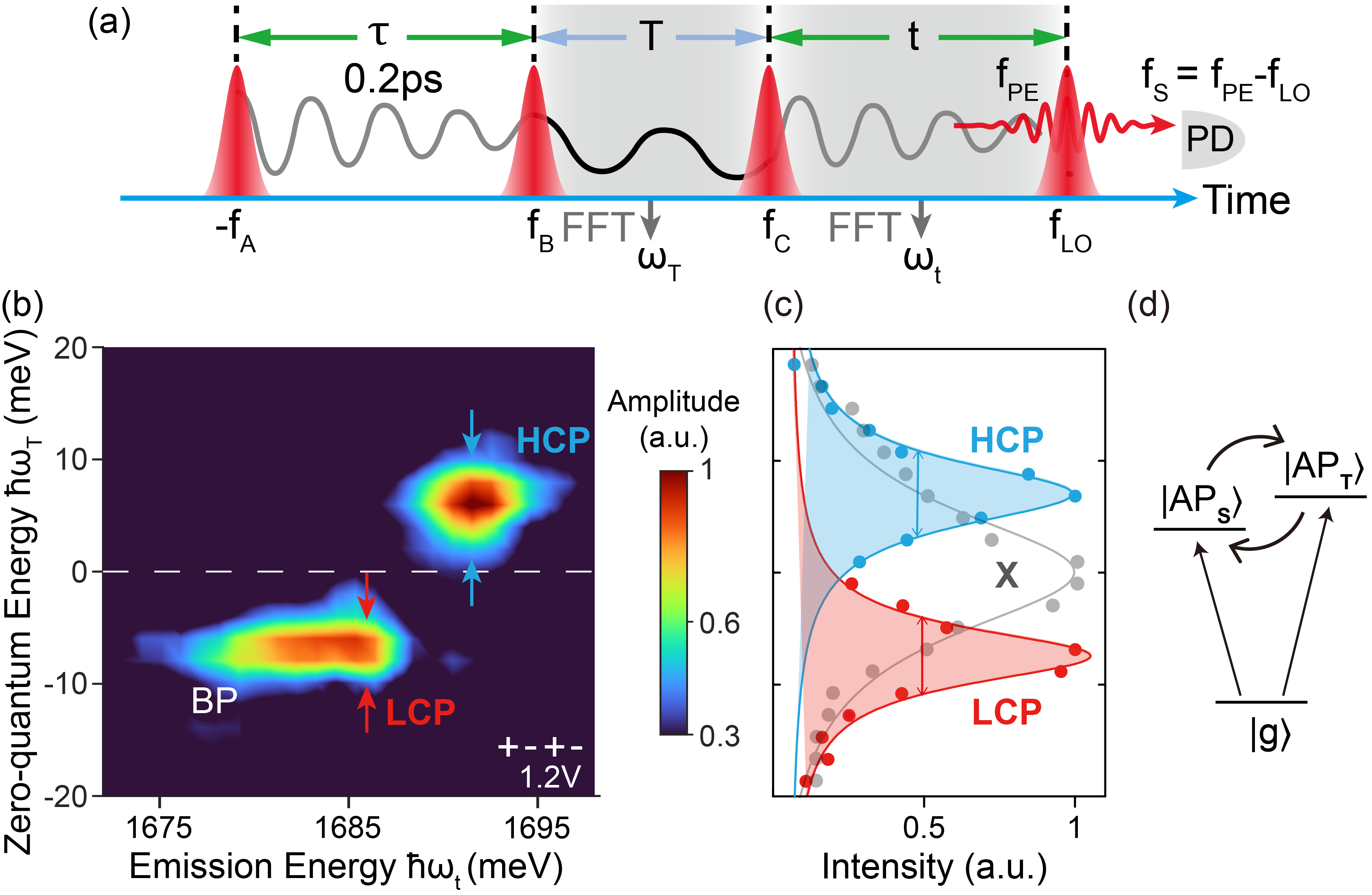}
    \caption{Zero-quantum spectra and valley coherence. 
    (a) The pulse sequence used in taking the zero-quantum spectra where delays $T$ and $t$ are scanned and Fourier transformed to obtain the zero-quantum and emission frequency axes ($\hbar\omega_T$ and $\hbar\omega_t$).
    (b) Normalized zero-quantum amplitude spectra under (+$-$+$-$) polarization at gate voltages \unit[1.2]{V}. 
    (c) Lorentzian fits of the vertical line-cut at energies of AP$_{S}$, AP$_{T}$ indicated by arrows in b, and exciton X (2D spectrum included in SI~\cite{SI}).
    (d) The level diagram showing the non-radiative valley coherence between AP$_{S}$ and AP$_{T}$ coupling to the same Fermi sea.
    }
    \label{fig3}
\end{figure}

We evaluate the non-radiative valley coherence between AP$_S$ and AP$_T$ in opposite valleys by taking zero-quantum spectra with alternating cross-circular polarization (+$-$+$-$).
Here, the time delays $T$ and $t$ are scanned and subsequently Fourier transformed to obtain two frequency axes as illustrated in Fig.~\ref{fig3}a. The delay between the first two pulses $\tau$ is fixed at \unit[0.2]{ps} to avoid coherent artifact terms. 
The first two cross-circularly polarized pulses create attractive polarons involving electron-hole pairs in opposite valleys. Because singlet and triplet states created in opposite valleys share the same Fermi sea, their interaction leads to a higher energy cross peak and a lower energy cross peak (HCP and LCP) as observed in Fig.~\ref{fig3}b. The non-radiative coherence between AP$_{S}$ and AP$_{T}$ evolves during the time delay $T$, yielding a shift of the two peaks (HCP and LCP) from $\omega_{T} = 0$ by $\sim$\unit[7]{meV}, the energy splitting between the AP$_{S}$ and AP$_{T}$ resonances, as expected. The elongation on the lower energy side of the LCP depicted in Fig.~\ref{fig3}b is attributed to bound bipolaron (BP) states similar to a state reported in WS$_{2}$~\cite{muir_interactions_2022} in one-quantum spectrum. Despite the shared optical selection rule with the biexciton, we suggest that the bipolaron should be considered as a bound state between two polarons because its doping density-dependent linedwidth follows a similar trend as polarons (details in SI~\cite{SI}).

We show the energy level diagram corresponding to the non-radiative valley coherence in Fig.~\ref{fig3}d, where the ground state refers to that of the doped semiconductor without any optical excitation. The non-radiative coherence of AP$_{S}$ and AP$_{T}$ manifests itself in the second order of the perturbation theory describing the light-matter interaction (more details in SI). In the simplest picture, the analysis of these two peaks should lead to the same non-radiative decoherence rate. The average of the HWHM linewidths from Lorentzian fits to the vertical line cuts of the HCP (blue) and LCP (red) (Fig.~\ref{fig3}c) is $\gamma_v=\unit[3.2]{meV}$, corresponding to a non-radiative coherence time of $\sim$~\unit[200]{fs}, faster than the predicted valley coherence time $\sim$~\unit[0.7]{ps} limited by the population relaxation. Thus, we conclude that this valley decoherence is dominated by pure dephasing processes~\cite{SI}. This attractive polaron valley coherence persists longer than exciton valley coherence (grey dots with a fitting curve as solid line) $\sim$~\unit[100]{fs}, which is surprising given that the attractive polaron dephasing times are comparable to that of the excitons. 

The pure polaron valley dephasing rate is related to other dephasing processes via 
\begin{equation}
    \gamma_v^*=\gamma^{*}_S+\gamma^{*}_T-2\cdot R\cdot\sqrt{\gamma^{*}_S \gamma^{*}_T}
\end{equation}
where $\gamma^{*}_S$ and $\gamma^{*}_T$ are the pure dephasing rates of AP$_S$ and AP$_T$, respectively. The coefficient $R$ quantifies the level of correlation between the energy fluctuations of the AP$_S$ and AP$_T$. Its possible values range from -1 to 1 with $R$ = 0 representing completely uncorrelated energy fluctuations. We extract a positive correlation coefficient $R = 0.22$ for their transition energy fluctuations (see details in the SI~\cite{SI}). For comparison, dephasing of excitons and trions in monolayer MoSe$_2$ was found to be uncorrelated \cite{2016_H.Kai_NanoLett_CP-2DCS}, while dephasing of light-hole and heavy-hole excitons in GaAs quantum wells was found to even be anti-correlated \cite{2008_VanEngen_GaAsCorrelationCoefficient}. We attribute this surprising correlated behavior to the unique coupling mechanism between AP$_S$ and AP$_T$ via a shared Fermi sea. In other systems, the non-radiative decoherence may be dominated by coupling between the electronic transitions and the phonon bath.

\begin{figure}
    \centering
    \includegraphics[width=8.6cm]{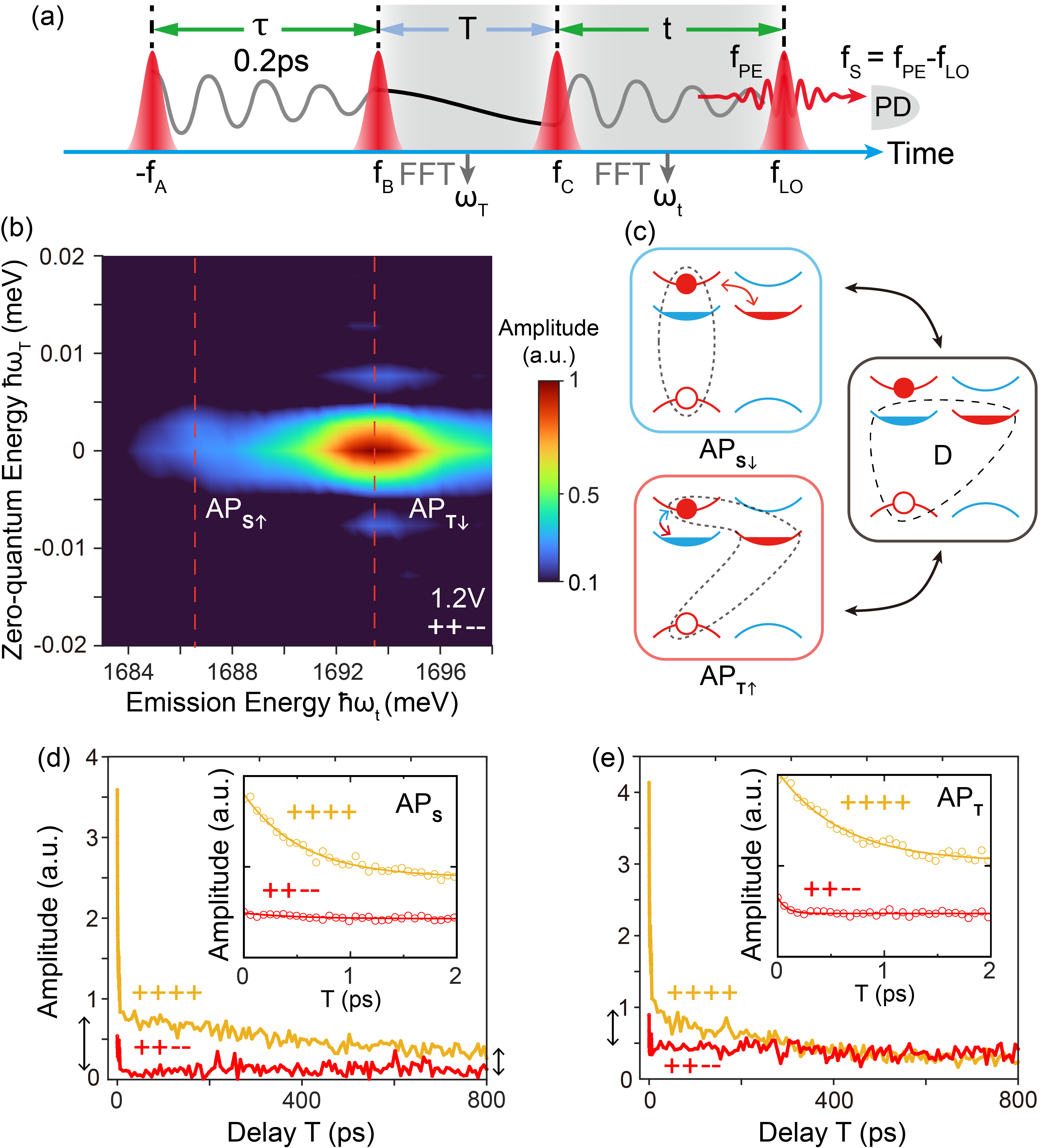}
    \caption{Zero-quantum spectra revealing a long-lived valley polarization of AP$_S$ and AP$_T$.
    (a) The pulse sequence used in taking the zero-quantum spectra, where the population decay or valley relaxation occurs during the delay $T$.
    (b) Normalized zero-quantum amplitude spectrum with $T$ scanned for \unit[800]{ps} at \unit[1.2]{V} taken with the polarization of ++$-$$-$.
    (c) Dark-to-bright state conversion contributes to the long-lived valley polarization of AP$_{S}$ and AP$_{T}$. Possible processes leading to dark states in WX$_2$. The enclosed areas indicate the conduction-band electrons and valence-band holes involved in the attractive polaron. The double-headed blue or red arrows indicate the electrons that are swapped between the bright and dark attractive polaron configurations.
    (d) and (e) Relaxation dynamics in the time domain for AP$_{S}$ and AP$_{T}$, respectively, extracted at energies indicated by vertical dashed lines in (b). The yellow (red) color corresponds to the population (valley polarization) relaxation dynamics with the gaps indicated by double-head black arrows, and the fast decays within the initial \unit[2]{ps} are displayed as insets. Additional spectra taken with ++++ polarization are included in the SI~\cite{SI}. 
    }
    \label{fig4}
\end{figure}

Although quantum dephasing of attractive polarons occurs on an ultrafast sub-picosecond time scale, consistent with their large oscillator strength, we observe a surprisingly long-lived valley polarization component associated with both attractive polarons by taking zero-quantum spectra with (++$-$$-$) polarization configuration, as illustrated in Fig.~\ref{fig4}a. This valley polarization persists over a few hundred picoseconds as evidenced by the ultranarrow line width for both AP$_{S}$ and AP$_{T}$ along the vertical direction (i.e., $\omega_{T}$) in Fig.~\ref{fig4}b, approximately 1000 times narrower than the linewidth reported in a 2DCS previous study on WS$_2$~\cite{muir_interactions_2022}.  
Converting this spectral feature to the time domain, 
we observe a rapid initial decay (insets) followed by long-lived population (++++) and valley polarization (++$-$$-$) components in Fig.~\ref{fig4}d-e.

Since the long-lived population and valley relaxation components are only observed in WSe$_2$ monolayers but not in MoSe$_2$ monolayers \cite{rodek_interactions_2024} (shown in the SI~\cite{SI}), we suggest that they originate from dark-to-bright state conversion processes that are unique to TMDs with inverted conduction bands.
Several different types of dark states are known to exist in WX$_2$ monolayers, including momentum-indirect transitions, spin forbidden states, and defect bound states~\cite{zhang_experimental_2015,li_momentum-dark_2019,robert_fine_2017,molas_probing_2019,zhou_probing_2017,li_direct_2019}. A likely candidate for the dark state configuration involved in our experiment is shown in Fig.~\ref{fig4}c: 
Here, the attractive polaron AP$_{S \downarrow}$ (AP$_{T \uparrow}$) has swapped its electron from the upper $\downarrow$ ($\uparrow$) conduction band with an electron from the otherwise inert $\downarrow$ ($\uparrow$) Fermi sea in the lower conduction band.
Crucially, due to the band inversion in WSe$_2$, the resulting configuration cannot undergo valley relaxation and it is thus expected to be long lived. This dark state is qualitatively distinct from the case of attractive polarons in MoSe$_2$, in which all electrons involved in the polaron state already reside in the lowest of the spin-split conduction bands and valley relaxation is possible~\cite{SI}. Furthermore, the WSe$_2$ dark-to-bright state conversion process is incoherent, thus only impacting the valley relaxation and not the valley dephasing processes.
Another prominent feature in Fig.~\ref{fig4}b is that the AP$_{S\uparrow}$ intensity is much weaker than that of AP$_{T\downarrow}$. 
This is consistent with the fact that AP$_S$ and AP$_T$ exhibit different valley relaxation timescales in Fig.~\ref{fig4}d-e. While the population decays of singlet and triplet attractive polarons measured using (++++) polarization configuration are similar, AP$_S$ has a lower valley scattering rate indicated by the smaller signal under (++$-$$-$) polarization. The different AP$_S$ and  AP$_T$ valley scattering processes are illustrated in Fig.~\ref{fig4}c. The electron-electron scattering process that links AP$_S$ to the dark state is affected by exchange interactions, while this is not the case for AP$_T$ where the switched electrons have different spins.

In conclusion, we have shown that the quantum dynamics of singlet and triplet attractive polarons in a WSe$_2$ monolayer systematically evolve with electron doping density. While dephasing rates of attractive polarons are stable at low doping density, they increase at higher doping density likely due to additional decay channels via coupling to higher-order many-body states.
The non-radiative valley coherence between the singlet- and triplet polarons persists longer than that of excitons, and its dephasing is dominated by pure dephasing processes rather than population relaxations. Analysis of valley and polaron pure dephasing processes reveals correlated energy fluctuations between the coupled singlet- and triplet states sharing the same Fermi sea. Remarkably, we have discovered a long-lived attractive polaron valley polarization component originating from a dark-to-bright state conversion. 
Notably, the scattering to and from the dark state is different for the singlet and the triplet polarons 
since there are electron-electron exchange interactions between the singlet configuration and the Fermi sea in the opposite valley. As a result, the singlet state exhibits a longer-lived valley polarization than that of the triplet state. Our studies identify crucial similarities and distinctions of polaron dynamics among different TMD monolayers and set the stage for exploring rich polaron phenomena predicted for moir\'e superlattices consisting of stacked TMD bilayers.


\section*{Acknowledgments}
The spectroscopic experiments performed by N. Y., D. H., D. L., K.S. and X. Liu at UT-Austin were primarily supported by the Department of Energy, Basic Energy Science program via grant DE-SC0019398. X. Liu and Z. Liu are partially supported by NSF MRSEC under Cooperative Agreement NO. DMR-2308817. Z. Liu acknowledges support from NSF ECCS-2130552. The work was partly done at the Texas Nanofabrication Facility supported by NSF grant NNCI-2025227. X.L. gratefully acknowledges sample preparation support by the Welch Foundation Chair F-0014. J.L., D.K.E., and M.M.P. acknowledge support from the Australian Research Council (ARC) Centre of Excellence in Future Low-Energy Electronics Technologies (CE170100039). J.L. and M.M.P. are also supported through the ARC Future Fellowships FT160100244 and FT200100619, respectively, and J.L., D.K.E., and X.L. furthermore acknowledge support from the ARC Discovery Project DP240100569. K.W. and T.T. acknowledge support from the JSPS KAKENHI (Grant Numbers 20H00354 and 23H02052) and World Premier International Research Center Initiative (WPI), MEXT, Japan.
N.Y. acknowledges device fabrication support by Dr. Yuanbo Zhang from Fudan University.
We thank Dr. Jeffrey A Davis for the technical discussions and suggestions.


\bibliography{wse2} 

\begin{thebibliography}{41}%
\makeatletter
\providecommand \@ifxundefined [1]{%
 \@ifx{#1\undefined}
}%
\providecommand \@ifnum [1]{%
 \ifnum #1\expandafter \@firstoftwo
 \else \expandafter \@secondoftwo
 \fi
}%
\providecommand \@ifx [1]{%
 \ifx #1\expandafter \@firstoftwo
 \else \expandafter \@secondoftwo
 \fi
}%
\providecommand \natexlab [1]{#1}%
\providecommand \enquote  [1]{``#1''}%
\providecommand \bibnamefont  [1]{#1}%
\providecommand \bibfnamefont [1]{#1}%
\providecommand \citenamefont [1]{#1}%
\providecommand \href@noop [0]{\@secondoftwo}%
\providecommand \href [0]{\begingroup \@sanitize@url \@href}%
\providecommand \@href[1]{\@@startlink{#1}\@@href}%
\providecommand \@@href[1]{\endgroup#1\@@endlink}%
\providecommand \@sanitize@url [0]{\catcode `\\12\catcode `\$12\catcode `\&12\catcode `\#12\catcode `\^12\catcode `\_12\catcode `\%12\relax}%
\providecommand \@@startlink[1]{}%
\providecommand \@@endlink[0]{}%
\providecommand \url  [0]{\begingroup\@sanitize@url \@url }%
\providecommand \@url [1]{\endgroup\@href {#1}{\urlprefix }}%
\providecommand \urlprefix  [0]{URL }%
\providecommand \Eprint [0]{\href }%
\providecommand \doibase [0]{http://dx.doi.org/}%
\providecommand \selectlanguage [0]{\@gobble}%
\providecommand \bibinfo  [0]{\@secondoftwo}%
\providecommand \bibfield  [0]{\@secondoftwo}%
\providecommand \translation [1]{[#1]}%
\providecommand \BibitemOpen [0]{}%
\providecommand \bibitemStop [0]{}%
\providecommand \bibitemNoStop [0]{.\EOS\space}%
\providecommand \EOS [0]{\spacefactor3000\relax}%
\providecommand \BibitemShut  [1]{\csname bibitem#1\endcsname}%
\let\auto@bib@innerbib\@empty
\bibitem [{\citenamefont {Massignan}\ \emph {et~al.}(2014)\citenamefont {Massignan}, \citenamefont {Zaccanti},\ and\ \citenamefont {Bruun}}]{2014_Massignan_RepProgPhys_Review}%
  \BibitemOpen
  \bibfield  {author} {\bibinfo {author} {\bibfnamefont {P.}~\bibnamefont {Massignan}}, \bibinfo {author} {\bibfnamefont {M.}~\bibnamefont {Zaccanti}}, \ and\ \bibinfo {author} {\bibfnamefont {G.~M.}\ \bibnamefont {Bruun}},\ }\href {\doibase 10.1088/0034-4885/77/3/034401} {\bibfield  {journal} {\bibinfo  {journal} {Rep Prog Phys}\ }\textbf {\bibinfo {volume} {77}},\ \bibinfo {pages} {034401} (\bibinfo {year} {2014})}\BibitemShut {NoStop}%
\bibitem [{\citenamefont {Levinsen}\ and\ \citenamefont {Parish}()}]{FermiGasesReview}%
  \BibitemOpen
  \bibfield  {author} {\bibinfo {author} {\bibfnamefont {J.}~\bibnamefont {Levinsen}}\ and\ \bibinfo {author} {\bibfnamefont {M.~M.}\ \bibnamefont {Parish}},\ }\enquote {\bibinfo {title} {Strongly interacting two-dimensional fermi gases},}\ in\ \href {\doibase 10.1142/9789814667746_0001} {\emph {\bibinfo {booktitle} {Annual Review of Cold Atoms and Molecules}}},\ Chap.~\bibinfo {chapter} {1}, pp.\ \bibinfo {pages} {1--75}\BibitemShut {NoStop}%
\bibitem [{\citenamefont {Sidler}\ \emph {et~al.}(2017)\citenamefont {Sidler}, \citenamefont {Back}, \citenamefont {Cotlet}, \citenamefont {Srivastava}, \citenamefont {Fink}, \citenamefont {Kroner}, \citenamefont {Demler},\ and\ \citenamefont {Imamoglu}}]{2016_A.Imamoglu_NatPhys_MoSe2}%
  \BibitemOpen
  \bibfield  {author} {\bibinfo {author} {\bibfnamefont {M.}~\bibnamefont {Sidler}}, \bibinfo {author} {\bibfnamefont {P.}~\bibnamefont {Back}}, \bibinfo {author} {\bibfnamefont {O.}~\bibnamefont {Cotlet}}, \bibinfo {author} {\bibfnamefont {A.}~\bibnamefont {Srivastava}}, \bibinfo {author} {\bibfnamefont {T.}~\bibnamefont {Fink}}, \bibinfo {author} {\bibfnamefont {M.}~\bibnamefont {Kroner}}, \bibinfo {author} {\bibfnamefont {E.}~\bibnamefont {Demler}}, \ and\ \bibinfo {author} {\bibfnamefont {A.}~\bibnamefont {Imamoglu}},\ }\href {\doibase 10.1038/nphys3949} {\bibfield  {journal} {\bibinfo  {journal} {Nature Physics}\ }\textbf {\bibinfo {volume} {13}},\ \bibinfo {pages} {255} (\bibinfo {year} {2017})}\BibitemShut {NoStop}%
\bibitem [{\citenamefont {Efimkin}\ and\ \citenamefont {MacDonald}(2017)}]{2017_Dmitry_PRB_theory}%
  \BibitemOpen
  \bibfield  {author} {\bibinfo {author} {\bibfnamefont {D.~K.}\ \bibnamefont {Efimkin}}\ and\ \bibinfo {author} {\bibfnamefont {A.~H.}\ \bibnamefont {MacDonald}},\ }\href {\doibase 10.1103/physrevb.95.035417} {\bibfield  {journal} {\bibinfo  {journal} {Physical Review B}\ }\textbf {\bibinfo {volume} {95}},\ \bibinfo {pages} {035417} (\bibinfo {year} {2017})}\BibitemShut {NoStop}%
\bibitem [{\citenamefont {Efimkin}\ and\ \citenamefont {MacDonald}(2018)}]{2019_Dmitry_PRB_theory}%
  \BibitemOpen
  \bibfield  {author} {\bibinfo {author} {\bibfnamefont {D.~K.}\ \bibnamefont {Efimkin}}\ and\ \bibinfo {author} {\bibfnamefont {A.~H.}\ \bibnamefont {MacDonald}},\ }\href {\doibase 10.1103/PhysRevB.97.235432} {\bibfield  {journal} {\bibinfo  {journal} {Phys. Rev. B}\ }\textbf {\bibinfo {volume} {97}},\ \bibinfo {pages} {235432} (\bibinfo {year} {2018})}\BibitemShut {NoStop}%
\bibitem [{\citenamefont {Muir}\ \emph {et~al.}(2022)\citenamefont {Muir}, \citenamefont {Levinsen}, \citenamefont {Earl}, \citenamefont {Conway}, \citenamefont {Cole}, \citenamefont {Wurdack}, \citenamefont {Mishra}, \citenamefont {Ing}, \citenamefont {Estrecho}, \citenamefont {Lu}, \citenamefont {Efimkin}, \citenamefont {Tollerud}, \citenamefont {Ostrovskaya}, \citenamefont {Parish},\ and\ \citenamefont {Davis}}]{muir_interactions_2022}%
  \BibitemOpen
  \bibfield  {author} {\bibinfo {author} {\bibfnamefont {J.~B.}\ \bibnamefont {Muir}}, \bibinfo {author} {\bibfnamefont {J.}~\bibnamefont {Levinsen}}, \bibinfo {author} {\bibfnamefont {S.~K.}\ \bibnamefont {Earl}}, \bibinfo {author} {\bibfnamefont {M.~A.}\ \bibnamefont {Conway}}, \bibinfo {author} {\bibfnamefont {J.~H.}\ \bibnamefont {Cole}}, \bibinfo {author} {\bibfnamefont {M.}~\bibnamefont {Wurdack}}, \bibinfo {author} {\bibfnamefont {R.}~\bibnamefont {Mishra}}, \bibinfo {author} {\bibfnamefont {D.~J.}\ \bibnamefont {Ing}}, \bibinfo {author} {\bibfnamefont {E.}~\bibnamefont {Estrecho}}, \bibinfo {author} {\bibfnamefont {Y.}~\bibnamefont {Lu}}, \bibinfo {author} {\bibfnamefont {D.~K.}\ \bibnamefont {Efimkin}}, \bibinfo {author} {\bibfnamefont {J.~O.}\ \bibnamefont {Tollerud}}, \bibinfo {author} {\bibfnamefont {E.~A.}\ \bibnamefont {Ostrovskaya}}, \bibinfo {author} {\bibfnamefont {M.~M.}\ \bibnamefont {Parish}}, \ and\ \bibinfo {author} {\bibfnamefont {J.~A.}\ \bibnamefont {Davis}},\ }\href {\doibase
  10.1038/s41467-022-33811-x} {\bibfield  {journal} {\bibinfo  {journal} {Nature Communications}\ }\textbf {\bibinfo {volume} {13}},\ \bibinfo {pages} {6164} (\bibinfo {year} {2022})}\BibitemShut {NoStop}%
\bibitem [{\citenamefont {Huang}\ \emph {et~al.}(2023)\citenamefont {Huang}, \citenamefont {Sampson}, \citenamefont {Ni}, \citenamefont {Liu}, \citenamefont {Liang}, \citenamefont {Watanabe}, \citenamefont {Taniguchi}, \citenamefont {Li}, \citenamefont {Martin}, \citenamefont {Levinsen}, \citenamefont {Parish}, \citenamefont {Tutuc}, \citenamefont {Efimkin},\ and\ \citenamefont {Li}}]{huang_quantum_2023}%
  \BibitemOpen
  \bibfield  {author} {\bibinfo {author} {\bibfnamefont {D.}~\bibnamefont {Huang}}, \bibinfo {author} {\bibfnamefont {K.}~\bibnamefont {Sampson}}, \bibinfo {author} {\bibfnamefont {Y.}~\bibnamefont {Ni}}, \bibinfo {author} {\bibfnamefont {Z.}~\bibnamefont {Liu}}, \bibinfo {author} {\bibfnamefont {D.}~\bibnamefont {Liang}}, \bibinfo {author} {\bibfnamefont {K.}~\bibnamefont {Watanabe}}, \bibinfo {author} {\bibfnamefont {T.}~\bibnamefont {Taniguchi}}, \bibinfo {author} {\bibfnamefont {H.}~\bibnamefont {Li}}, \bibinfo {author} {\bibfnamefont {E.}~\bibnamefont {Martin}}, \bibinfo {author} {\bibfnamefont {J.}~\bibnamefont {Levinsen}}, \bibinfo {author} {\bibfnamefont {M.~M.}\ \bibnamefont {Parish}}, \bibinfo {author} {\bibfnamefont {E.}~\bibnamefont {Tutuc}}, \bibinfo {author} {\bibfnamefont {D.~K.}\ \bibnamefont {Efimkin}}, \ and\ \bibinfo {author} {\bibfnamefont {X.}~\bibnamefont {Li}},\ }\href {\doibase 10.1103/PhysRevX.13.011029} {\bibfield  {journal} {\bibinfo  {journal} {Physical Review X}\ }\textbf
  {\bibinfo {volume} {13}},\ \bibinfo {pages} {011029} (\bibinfo {year} {2023})}\BibitemShut {NoStop}%
\bibitem [{\citenamefont {Wang}\ \emph {et~al.}(2018)\citenamefont {Wang}, \citenamefont {Chernikov}, \citenamefont {Glazov}, \citenamefont {Heinz}, \citenamefont {Marie}, \citenamefont {Amand},\ and\ \citenamefont {Urbaszek}}]{2018_G.Wang_RMP_ExcitonReview}%
  \BibitemOpen
  \bibfield  {author} {\bibinfo {author} {\bibfnamefont {G.}~\bibnamefont {Wang}}, \bibinfo {author} {\bibfnamefont {A.}~\bibnamefont {Chernikov}}, \bibinfo {author} {\bibfnamefont {M.~M.}\ \bibnamefont {Glazov}}, \bibinfo {author} {\bibfnamefont {T.~F.}\ \bibnamefont {Heinz}}, \bibinfo {author} {\bibfnamefont {X.}~\bibnamefont {Marie}}, \bibinfo {author} {\bibfnamefont {T.}~\bibnamefont {Amand}}, \ and\ \bibinfo {author} {\bibfnamefont {B.}~\bibnamefont {Urbaszek}},\ }\href {\doibase 10.1103/revmodphys.90.021001} {\bibfield  {journal} {\bibinfo  {journal} {Reviews of Modern Physics}\ }\textbf {\bibinfo {volume} {90}},\ \bibinfo {pages} {021001} (\bibinfo {year} {2018})}\BibitemShut {NoStop}%
\bibitem [{\citenamefont {Xiao}\ \emph {et~al.}(2012)\citenamefont {Xiao}, \citenamefont {Liu}, \citenamefont {Feng}, \citenamefont {Xu},\ and\ \citenamefont {Yao}}]{xiao_coupled_2012}%
  \BibitemOpen
  \bibfield  {author} {\bibinfo {author} {\bibfnamefont {D.}~\bibnamefont {Xiao}}, \bibinfo {author} {\bibfnamefont {G.-B.}\ \bibnamefont {Liu}}, \bibinfo {author} {\bibfnamefont {W.}~\bibnamefont {Feng}}, \bibinfo {author} {\bibfnamefont {X.}~\bibnamefont {Xu}}, \ and\ \bibinfo {author} {\bibfnamefont {W.}~\bibnamefont {Yao}},\ }\href {\doibase 10.1103/PhysRevLett.108.196802} {\bibfield  {journal} {\bibinfo  {journal} {Physical Review Letters}\ }\textbf {\bibinfo {volume} {108}},\ \bibinfo {pages} {196802} (\bibinfo {year} {2012})}\BibitemShut {NoStop}%
\bibitem [{\citenamefont {Vaclavkova}\ \emph {et~al.}(2018)\citenamefont {Vaclavkova}, \citenamefont {Wyzula}, \citenamefont {Nogajewski}, \citenamefont {Bartos}, \citenamefont {Slobodeniuk}, \citenamefont {Faugeras}, \citenamefont {Potemski},\ and\ \citenamefont {Molas}}]{Vaclavkova_2018}%
  \BibitemOpen
  \bibfield  {author} {\bibinfo {author} {\bibfnamefont {D.}~\bibnamefont {Vaclavkova}}, \bibinfo {author} {\bibfnamefont {J.}~\bibnamefont {Wyzula}}, \bibinfo {author} {\bibfnamefont {K.}~\bibnamefont {Nogajewski}}, \bibinfo {author} {\bibfnamefont {M.}~\bibnamefont {Bartos}}, \bibinfo {author} {\bibfnamefont {A.~O.}\ \bibnamefont {Slobodeniuk}}, \bibinfo {author} {\bibfnamefont {C.}~\bibnamefont {Faugeras}}, \bibinfo {author} {\bibfnamefont {M.}~\bibnamefont {Potemski}}, \ and\ \bibinfo {author} {\bibfnamefont {M.~R.}\ \bibnamefont {Molas}},\ }\href {\doibase 10.1088/1361-6528/aac65c} {\bibfield  {journal} {\bibinfo  {journal} {Nanotechnology}\ }\textbf {\bibinfo {volume} {29}},\ \bibinfo {pages} {325705} (\bibinfo {year} {2018})}\BibitemShut {NoStop}%
\bibitem [{\citenamefont {Jadczak}\ \emph {et~al.}(2021)\citenamefont {Jadczak}, \citenamefont {Glazov}, \citenamefont {Kutrowska-Girzycka}, \citenamefont {Schindler}, \citenamefont {Debus}, \citenamefont {Ho}, \citenamefont {Watanabe}, \citenamefont {Taniguchi}, \citenamefont {Bayer},\ and\ \citenamefont {Bryja}}]{jadczak_upconversion_2021}%
  \BibitemOpen
  \bibfield  {author} {\bibinfo {author} {\bibfnamefont {J.}~\bibnamefont {Jadczak}}, \bibinfo {author} {\bibfnamefont {M.}~\bibnamefont {Glazov}}, \bibinfo {author} {\bibfnamefont {J.}~\bibnamefont {Kutrowska-Girzycka}}, \bibinfo {author} {\bibfnamefont {J.~J.}\ \bibnamefont {Schindler}}, \bibinfo {author} {\bibfnamefont {J.}~\bibnamefont {Debus}}, \bibinfo {author} {\bibfnamefont {C.-H.}\ \bibnamefont {Ho}}, \bibinfo {author} {\bibfnamefont {K.}~\bibnamefont {Watanabe}}, \bibinfo {author} {\bibfnamefont {T.}~\bibnamefont {Taniguchi}}, \bibinfo {author} {\bibfnamefont {M.}~\bibnamefont {Bayer}}, \ and\ \bibinfo {author} {\bibfnamefont {L.}~\bibnamefont {Bryja}},\ }\href {\doibase 10.1021/acsnano.1c08286} {\bibfield  {journal} {\bibinfo  {journal} {ACS Nano}\ }\textbf {\bibinfo {volume} {15}},\ \bibinfo {pages} {19165} (\bibinfo {year} {2021})}\BibitemShut {NoStop}%
\bibitem [{\citenamefont {Danovich}\ \emph {et~al.}(2017)\citenamefont {Danovich}, \citenamefont {Zólyomi},\ and\ \citenamefont {Fal’ko}}]{danovich_dark_2017}%
  \BibitemOpen
  \bibfield  {author} {\bibinfo {author} {\bibfnamefont {M.}~\bibnamefont {Danovich}}, \bibinfo {author} {\bibfnamefont {V.}~\bibnamefont {Zólyomi}}, \ and\ \bibinfo {author} {\bibfnamefont {V.~I.}\ \bibnamefont {Fal’ko}},\ }\href {\doibase 10.1038/srep45998} {\bibfield  {journal} {\bibinfo  {journal} {Sci Rep}\ }\textbf {\bibinfo {volume} {7}},\ \bibinfo {pages} {45998} (\bibinfo {year} {2017})}\BibitemShut {NoStop}%
\bibitem [{\citenamefont {Cong}\ \emph {et~al.}(2023)\citenamefont {Cong}, \citenamefont {Mohammadi}, \citenamefont {Zheng}, \citenamefont {Watanabe}, \citenamefont {Taniguchi}, \citenamefont {Rhodes},\ and\ \citenamefont {Zhang}}]{cong_interplay_2023}%
  \BibitemOpen
  \bibfield  {author} {\bibinfo {author} {\bibfnamefont {X.}~\bibnamefont {Cong}}, \bibinfo {author} {\bibfnamefont {P.~A.}\ \bibnamefont {Mohammadi}}, \bibinfo {author} {\bibfnamefont {M.}~\bibnamefont {Zheng}}, \bibinfo {author} {\bibfnamefont {K.}~\bibnamefont {Watanabe}}, \bibinfo {author} {\bibfnamefont {T.}~\bibnamefont {Taniguchi}}, \bibinfo {author} {\bibfnamefont {D.}~\bibnamefont {Rhodes}}, \ and\ \bibinfo {author} {\bibfnamefont {X.-X.}\ \bibnamefont {Zhang}},\ }\href {\doibase 10.1038/s41467-023-41475-4} {\bibfield  {journal} {\bibinfo  {journal} {Nat Commun}\ }\textbf {\bibinfo {volume} {14}},\ \bibinfo {pages} {5657} (\bibinfo {year} {2023})}\BibitemShut {NoStop}%
\bibitem [{see Supplemental Material at [URL], which includes Refs. [15-19], for additional information about (A) 2DCES experiment, (B) Fabrication of the monolayer device, (C) Extracting $\Gamma$, $\gamma$, and $\gamma_{v}$ from one-quantum and zero-quantum 2DECS spectra respectively, (D) Fermi energy of the doped WSe$_2$ monolayer, (E) Evaluating oscillator strength from one-quantum spectra, (F) Quantum pathway of attractive polaron valley coherence, (G) Analysis of the valley coherence based on three-le)see Supplemental Material at [URL], which includes Refs. [15-19], for additional information about (A) 2DCES experiment, (B) Fabrication of the monolayer device, (C) Extracting $\Gamma$, $\gamma$, and $\gamma_{v}$ from one-quantum and zero-quantum 2DECS spectra respectively, (D) Fermi energy of the doped WSe$_2$ monolayer, (E) Evaluating oscillator strength from one-quantum spectra, (F) Quantum pathway of attractive polaron valley coherence, (G) Analysis of the valley coherence based on three-level
  V-system, (H) The Fermi polaron theory, (I) Analysis of bipolaron, (J) Ultrafast valley dynamics in MoSe2 monolayer, and (K) Original raw 2DCS spectra}]{SI}%
  \BibitemOpen
  see Supplemental Material at [URL], which includes Refs. [15-19], for additional information about (A) 2DCES experiment, (B) Fabrication of the monolayer device, (C) Extracting $\Gamma$, $\gamma$, and $\gamma_{v}$ from one-quantum and zero-quantum 2DECS spectra respectively, (D) Fermi energy of the doped WSe$_2$ monolayer, (E) Evaluating oscillator strength from one-quantum spectra, (F) Quantum pathway of attractive polaron valley coherence, (G) Analysis of the valley coherence based on three-level V-system, (H) The Fermi polaron theory, (I) Analysis of bipolaron, (J) Ultrafast valley dynamics in MoSe2 monolayer, and (K) Original raw 2DCS spectra,\ \href@noop {} {}\BibitemShut {NoStop}%
\bibitem [{\citenamefont {Nardin}\ \emph {et~al.}(2013)\citenamefont {Nardin}, \citenamefont {Autry}, \citenamefont {Silverman},\ and\ \citenamefont {Cundiff}}]{2013_G.Nardin_OptExp_2DCSSetup}%
  \BibitemOpen
  \bibfield  {author} {\bibinfo {author} {\bibfnamefont {G.}~\bibnamefont {Nardin}}, \bibinfo {author} {\bibfnamefont {T.~M.}\ \bibnamefont {Autry}}, \bibinfo {author} {\bibfnamefont {K.~L.}\ \bibnamefont {Silverman}}, \ and\ \bibinfo {author} {\bibfnamefont {S.~T.}\ \bibnamefont {Cundiff}},\ }\href {\doibase 10.1364/OE.21.028617} {\bibfield  {journal} {\bibinfo  {journal} {Opt Express}\ }\textbf {\bibinfo {volume} {21}},\ \bibinfo {pages} {28617} (\bibinfo {year} {2013})}\BibitemShut {NoStop}%
\bibitem [{\citenamefont {Moody}\ \emph {et~al.}(2015)\citenamefont {Moody}, \citenamefont {Kavir~Dass}, \citenamefont {Hao}, \citenamefont {Chen}, \citenamefont {Li}, \citenamefont {Singh}, \citenamefont {Tran}, \citenamefont {Clark}, \citenamefont {Xu}, \citenamefont {Berghauser}, \citenamefont {Malic}, \citenamefont {Knorr},\ and\ \citenamefont {Li}}]{2015_G.Moody_NatComm_2DCSMoSe2}%
  \BibitemOpen
  \bibfield  {author} {\bibinfo {author} {\bibfnamefont {G.}~\bibnamefont {Moody}}, \bibinfo {author} {\bibfnamefont {C.}~\bibnamefont {Kavir~Dass}}, \bibinfo {author} {\bibfnamefont {K.}~\bibnamefont {Hao}}, \bibinfo {author} {\bibfnamefont {C.~H.}\ \bibnamefont {Chen}}, \bibinfo {author} {\bibfnamefont {L.~J.}\ \bibnamefont {Li}}, \bibinfo {author} {\bibfnamefont {A.}~\bibnamefont {Singh}}, \bibinfo {author} {\bibfnamefont {K.}~\bibnamefont {Tran}}, \bibinfo {author} {\bibfnamefont {G.}~\bibnamefont {Clark}}, \bibinfo {author} {\bibfnamefont {X.}~\bibnamefont {Xu}}, \bibinfo {author} {\bibfnamefont {G.}~\bibnamefont {Berghauser}}, \bibinfo {author} {\bibfnamefont {E.}~\bibnamefont {Malic}}, \bibinfo {author} {\bibfnamefont {A.}~\bibnamefont {Knorr}}, \ and\ \bibinfo {author} {\bibfnamefont {X.}~\bibnamefont {Li}},\ }\href {\doibase 10.1038/ncomms9315} {\bibfield  {journal} {\bibinfo  {journal} {Nature Communications}\ }\textbf {\bibinfo {volume} {6}},\ \bibinfo {pages} {8315} (\bibinfo {year}
  {2015})}\BibitemShut {NoStop}%
\bibitem [{\citenamefont {Kormányos}\ \emph {et~al.}(2015)\citenamefont {Kormányos}, \citenamefont {Burkard}, \citenamefont {Gmitra}, \citenamefont {Fabian}, \citenamefont {Zólyomi}, \citenamefont {Drummond},\ and\ \citenamefont {Fal’Ko}}]{2015_Kormanyos_2DMat_MoSe2Review}%
  \BibitemOpen
  \bibfield  {author} {\bibinfo {author} {\bibfnamefont {A.}~\bibnamefont {Kormányos}}, \bibinfo {author} {\bibfnamefont {G.}~\bibnamefont {Burkard}}, \bibinfo {author} {\bibfnamefont {M.}~\bibnamefont {Gmitra}}, \bibinfo {author} {\bibfnamefont {J.}~\bibnamefont {Fabian}}, \bibinfo {author} {\bibfnamefont {V.}~\bibnamefont {Zólyomi}}, \bibinfo {author} {\bibfnamefont {N.~D.}\ \bibnamefont {Drummond}}, \ and\ \bibinfo {author} {\bibfnamefont {V.}~\bibnamefont {Fal’Ko}},\ }\href {\doibase 10.1088/2053-1583/2/2/022001} {\bibfield  {journal} {\bibinfo  {journal} {2D Materials}\ }\textbf {\bibinfo {volume} {2}},\ \bibinfo {pages} {022001} (\bibinfo {year} {2015})}\BibitemShut {NoStop}%
\bibitem [{\citenamefont {Yang}\ \emph {et~al.}(2020)\citenamefont {Yang}, \citenamefont {Robert}, \citenamefont {Lu}, \citenamefont {Van~Tuan}, \citenamefont {Smirnov}, \citenamefont {Marie},\ and\ \citenamefont {Dery}}]{D.Hanan_DBTheory_2020}%
  \BibitemOpen
  \bibfield  {author} {\bibinfo {author} {\bibfnamefont {M.}~\bibnamefont {Yang}}, \bibinfo {author} {\bibfnamefont {C.}~\bibnamefont {Robert}}, \bibinfo {author} {\bibfnamefont {Z.}~\bibnamefont {Lu}}, \bibinfo {author} {\bibfnamefont {D.}~\bibnamefont {Van~Tuan}}, \bibinfo {author} {\bibfnamefont {D.}~\bibnamefont {Smirnov}}, \bibinfo {author} {\bibfnamefont {X.}~\bibnamefont {Marie}}, \ and\ \bibinfo {author} {\bibfnamefont {H.}~\bibnamefont {Dery}},\ }\href {\doibase 10.1103/PhysRevB.101.115307} {\bibfield  {journal} {\bibinfo  {journal} {Physical Review B}\ }\textbf {\bibinfo {volume} {101}} (\bibinfo {year} {2020}),\ 10.1103/PhysRevB.101.115307}\BibitemShut {NoStop}%
\bibitem [{\citenamefont {Robert}\ \emph {et~al.}(2020)\citenamefont {Robert}, \citenamefont {Han}, \citenamefont {Kapuscinski}, \citenamefont {Delhomme}, \citenamefont {Faugeras}, \citenamefont {Amand}, \citenamefont {Molas}, \citenamefont {Bartos}, \citenamefont {Watanabe}, \citenamefont {Taniguchi}, \citenamefont {Urbaszek}, \citenamefont {Potemski},\ and\ \citenamefont {Marie}}]{robert_measurement_2020}%
  \BibitemOpen
  \bibfield  {author} {\bibinfo {author} {\bibfnamefont {C.}~\bibnamefont {Robert}}, \bibinfo {author} {\bibfnamefont {B.}~\bibnamefont {Han}}, \bibinfo {author} {\bibfnamefont {P.}~\bibnamefont {Kapuscinski}}, \bibinfo {author} {\bibfnamefont {A.}~\bibnamefont {Delhomme}}, \bibinfo {author} {\bibfnamefont {C.}~\bibnamefont {Faugeras}}, \bibinfo {author} {\bibfnamefont {T.}~\bibnamefont {Amand}}, \bibinfo {author} {\bibfnamefont {M.~R.}\ \bibnamefont {Molas}}, \bibinfo {author} {\bibfnamefont {M.}~\bibnamefont {Bartos}}, \bibinfo {author} {\bibfnamefont {K.}~\bibnamefont {Watanabe}}, \bibinfo {author} {\bibfnamefont {T.}~\bibnamefont {Taniguchi}}, \bibinfo {author} {\bibfnamefont {B.}~\bibnamefont {Urbaszek}}, \bibinfo {author} {\bibfnamefont {M.}~\bibnamefont {Potemski}}, \ and\ \bibinfo {author} {\bibfnamefont {X.}~\bibnamefont {Marie}},\ }\href {\doibase 10.1038/s41467-020-17608-4} {\bibfield  {journal} {\bibinfo  {journal} {Nat Commun}\ }\textbf {\bibinfo {volume} {11}},\ \bibinfo {pages} {4037} (\bibinfo
  {year} {2020})}\BibitemShut {NoStop}%
\bibitem [{\citenamefont {Glazov}(2020)}]{2020_Glazov_JCP_EqualTrionPolaron}%
  \BibitemOpen
  \bibfield  {author} {\bibinfo {author} {\bibfnamefont {M.~M.}\ \bibnamefont {Glazov}},\ }\href {\doibase 10.1063/5.0012475} {\bibfield  {journal} {\bibinfo  {journal} {The Journal of Chemical Physics}\ }\textbf {\bibinfo {volume} {153}},\ \bibinfo {pages} {034703} (\bibinfo {year} {2020})}\BibitemShut {NoStop}%
\bibitem [{\citenamefont {Suris}\ \emph {et~al.}(2001)\citenamefont {Suris}, \citenamefont {Kochereshko}, \citenamefont {Astakhov}, \citenamefont {Yakovlev}, \citenamefont {Ossau}, \citenamefont {Nürnberger}, \citenamefont {Faschinger}, \citenamefont {Landwehr}, \citenamefont {Wojtowicz}, \citenamefont {Karczewski},\ and\ \citenamefont {Kossut}}]{TrionDoubtSuris}%
  \BibitemOpen
  \bibfield  {author} {\bibinfo {author} {\bibfnamefont {R.}~\bibnamefont {Suris}}, \bibinfo {author} {\bibfnamefont {V.}~\bibnamefont {Kochereshko}}, \bibinfo {author} {\bibfnamefont {G.}~\bibnamefont {Astakhov}}, \bibinfo {author} {\bibfnamefont {D.}~\bibnamefont {Yakovlev}}, \bibinfo {author} {\bibfnamefont {W.}~\bibnamefont {Ossau}}, \bibinfo {author} {\bibfnamefont {J.}~\bibnamefont {Nürnberger}}, \bibinfo {author} {\bibfnamefont {W.}~\bibnamefont {Faschinger}}, \bibinfo {author} {\bibfnamefont {G.}~\bibnamefont {Landwehr}}, \bibinfo {author} {\bibfnamefont {T.}~\bibnamefont {Wojtowicz}}, \bibinfo {author} {\bibfnamefont {G.}~\bibnamefont {Karczewski}}, \ and\ \bibinfo {author} {\bibfnamefont {J.}~\bibnamefont {Kossut}},\ }\href {\doibase https://doi.org/10.1002/1521-3951(200110)227:2<343::AID-PSSB343>3.0.CO;2-W} {\bibfield  {journal} {\bibinfo  {journal} {physica status solidi (b)}\ }\textbf {\bibinfo {volume} {227}},\ \bibinfo {pages} {343} (\bibinfo {year} {2001})}\BibitemShut {NoStop}%
\bibitem [{\citenamefont {Baeten}\ and\ \citenamefont {Wouters}(2015)}]{TrionDoubtWouters2}%
  \BibitemOpen
  \bibfield  {author} {\bibinfo {author} {\bibfnamefont {M.}~\bibnamefont {Baeten}}\ and\ \bibinfo {author} {\bibfnamefont {M.}~\bibnamefont {Wouters}},\ }\href {\doibase 10.1103/PhysRevB.91.115313} {\bibfield  {journal} {\bibinfo  {journal} {Phys. Rev. B}\ }\textbf {\bibinfo {volume} {91}},\ \bibinfo {pages} {115313} (\bibinfo {year} {2015})}\BibitemShut {NoStop}%
\bibitem [{\citenamefont {Wagner}\ \emph {et~al.}(2020)\citenamefont {Wagner}, \citenamefont {Wietek}, \citenamefont {Ziegler}, \citenamefont {Semina}, \citenamefont {Taniguchi}, \citenamefont {Watanabe}, \citenamefont {Zipfel}, \citenamefont {Glazov},\ and\ \citenamefont {Chernikov}}]{2020_A.Chernikov_PRL_WSe2Trion}%
  \BibitemOpen
  \bibfield  {author} {\bibinfo {author} {\bibfnamefont {K.}~\bibnamefont {Wagner}}, \bibinfo {author} {\bibfnamefont {E.}~\bibnamefont {Wietek}}, \bibinfo {author} {\bibfnamefont {J.~D.}\ \bibnamefont {Ziegler}}, \bibinfo {author} {\bibfnamefont {M.~A.}\ \bibnamefont {Semina}}, \bibinfo {author} {\bibfnamefont {T.}~\bibnamefont {Taniguchi}}, \bibinfo {author} {\bibfnamefont {K.}~\bibnamefont {Watanabe}}, \bibinfo {author} {\bibfnamefont {J.}~\bibnamefont {Zipfel}}, \bibinfo {author} {\bibfnamefont {M.~M.}\ \bibnamefont {Glazov}}, \ and\ \bibinfo {author} {\bibfnamefont {A.}~\bibnamefont {Chernikov}},\ }\href {\doibase 10.1103/physrevlett.125.267401} {\bibfield  {journal} {\bibinfo  {journal} {Physical Review Letters}\ }\textbf {\bibinfo {volume} {125}},\ \bibinfo {pages} {267401} (\bibinfo {year} {2020})}\BibitemShut {NoStop}%
\bibitem [{\citenamefont {Liu}\ \emph {et~al.}(2021)\citenamefont {Liu}, \citenamefont {Van~Baren}, \citenamefont {Lu}, \citenamefont {Taniguchi}, \citenamefont {Watanabe}, \citenamefont {Smirnov}, \citenamefont {Chang},\ and\ \citenamefont {Lui}}]{2021_CHLui_NatComm_MoSe2&WSe2}%
  \BibitemOpen
  \bibfield  {author} {\bibinfo {author} {\bibfnamefont {E.}~\bibnamefont {Liu}}, \bibinfo {author} {\bibfnamefont {J.}~\bibnamefont {Van~Baren}}, \bibinfo {author} {\bibfnamefont {Z.}~\bibnamefont {Lu}}, \bibinfo {author} {\bibfnamefont {T.}~\bibnamefont {Taniguchi}}, \bibinfo {author} {\bibfnamefont {K.}~\bibnamefont {Watanabe}}, \bibinfo {author} {\bibfnamefont {D.}~\bibnamefont {Smirnov}}, \bibinfo {author} {\bibfnamefont {Y.-C.}\ \bibnamefont {Chang}}, \ and\ \bibinfo {author} {\bibfnamefont {C.~H.}\ \bibnamefont {Lui}},\ }\href {\doibase 10.1038/s41467-021-26304-w} {\bibfield  {journal} {\bibinfo  {journal} {Nature Communications}\ }\textbf {\bibinfo {volume} {12}} (\bibinfo {year} {2021}),\ 10.1038/s41467-021-26304-w}\BibitemShut {NoStop}%
\bibitem [{\citenamefont {Yu}\ \emph {et~al.}(2014)\citenamefont {Yu}, \citenamefont {Liu}, \citenamefont {Gong}, \citenamefont {Xu},\ and\ \citenamefont {Yao}}]{yu_dirac_2014}%
  \BibitemOpen
  \bibfield  {author} {\bibinfo {author} {\bibfnamefont {H.}~\bibnamefont {Yu}}, \bibinfo {author} {\bibfnamefont {G.-B.}\ \bibnamefont {Liu}}, \bibinfo {author} {\bibfnamefont {P.}~\bibnamefont {Gong}}, \bibinfo {author} {\bibfnamefont {X.}~\bibnamefont {Xu}}, \ and\ \bibinfo {author} {\bibfnamefont {W.}~\bibnamefont {Yao}},\ }\href {\doibase 10.1038/ncomms4876} {\bibfield  {journal} {\bibinfo  {journal} {Nature Communications}\ }\textbf {\bibinfo {volume} {5}},\ \bibinfo {pages} {3876} (\bibinfo {year} {2014})}\BibitemShut {NoStop}%
\bibitem [{\citenamefont {Lyons}\ \emph {et~al.}(2019)\citenamefont {Lyons}, \citenamefont {Dufferwiel}, \citenamefont {Brooks}, \citenamefont {Withers}, \citenamefont {Taniguchi}, \citenamefont {Watanabe}, \citenamefont {Novoselov}, \citenamefont {Burkard},\ and\ \citenamefont {Tartakovskii}}]{lyons_valley_2019}%
  \BibitemOpen
  \bibfield  {author} {\bibinfo {author} {\bibfnamefont {T.~P.}\ \bibnamefont {Lyons}}, \bibinfo {author} {\bibfnamefont {S.}~\bibnamefont {Dufferwiel}}, \bibinfo {author} {\bibfnamefont {M.}~\bibnamefont {Brooks}}, \bibinfo {author} {\bibfnamefont {F.}~\bibnamefont {Withers}}, \bibinfo {author} {\bibfnamefont {T.}~\bibnamefont {Taniguchi}}, \bibinfo {author} {\bibfnamefont {K.}~\bibnamefont {Watanabe}}, \bibinfo {author} {\bibfnamefont {K.~S.}\ \bibnamefont {Novoselov}}, \bibinfo {author} {\bibfnamefont {G.}~\bibnamefont {Burkard}}, \ and\ \bibinfo {author} {\bibfnamefont {A.~I.}\ \bibnamefont {Tartakovskii}},\ }\href {\doibase 10.1038/s41467-019-10228-7} {\bibfield  {journal} {\bibinfo  {journal} {Nature Communications}\ }\textbf {\bibinfo {volume} {10}},\ \bibinfo {pages} {2330} (\bibinfo {year} {2019})}\BibitemShut {NoStop}%
\bibitem [{\citenamefont {Van~Tuan}\ \emph {et~al.}(2022)\citenamefont {Van~Tuan}, \citenamefont {Shi}, \citenamefont {Xu}, \citenamefont {Crooker},\ and\ \citenamefont {Dery}}]{PhysRevLett.129.076801}%
  \BibitemOpen
  \bibfield  {author} {\bibinfo {author} {\bibfnamefont {D.}~\bibnamefont {Van~Tuan}}, \bibinfo {author} {\bibfnamefont {S.-F.}\ \bibnamefont {Shi}}, \bibinfo {author} {\bibfnamefont {X.}~\bibnamefont {Xu}}, \bibinfo {author} {\bibfnamefont {S.~A.}\ \bibnamefont {Crooker}}, \ and\ \bibinfo {author} {\bibfnamefont {H.}~\bibnamefont {Dery}},\ }\href {\doibase 10.1103/PhysRevLett.129.076801} {\bibfield  {journal} {\bibinfo  {journal} {Phys. Rev. Lett.}\ }\textbf {\bibinfo {volume} {129}},\ \bibinfo {pages} {076801} (\bibinfo {year} {2022})}\BibitemShut {NoStop}%
\bibitem [{\citenamefont {Nardin}\ \emph {et~al.}(2015)\citenamefont {Nardin}, \citenamefont {Autry}, \citenamefont {Moody}, \citenamefont {Singh}, \citenamefont {Li},\ and\ \citenamefont {Cundiff}}]{nardin_multi-dimensional_2015}%
  \BibitemOpen
  \bibfield  {author} {\bibinfo {author} {\bibfnamefont {G.}~\bibnamefont {Nardin}}, \bibinfo {author} {\bibfnamefont {T.~M.}\ \bibnamefont {Autry}}, \bibinfo {author} {\bibfnamefont {G.}~\bibnamefont {Moody}}, \bibinfo {author} {\bibfnamefont {R.}~\bibnamefont {Singh}}, \bibinfo {author} {\bibfnamefont {H.}~\bibnamefont {Li}}, \ and\ \bibinfo {author} {\bibfnamefont {S.~T.}\ \bibnamefont {Cundiff}},\ }\href {\doibase 10.1063/1.4913830} {\bibfield  {journal} {\bibinfo  {journal} {Journal of Applied Physics}\ }\textbf {\bibinfo {volume} {117}},\ \bibinfo {pages} {112804} (\bibinfo {year} {2015})}\BibitemShut {NoStop}%
\bibitem [{\citenamefont {Siemens}\ \emph {et~al.}(2010)\citenamefont {Siemens}, \citenamefont {Moody}, \citenamefont {Li}, \citenamefont {Bristow},\ and\ \citenamefont {Cundiff}}]{2010_Mark_OE_2DCSLineShape}%
  \BibitemOpen
  \bibfield  {author} {\bibinfo {author} {\bibfnamefont {M.~E.}\ \bibnamefont {Siemens}}, \bibinfo {author} {\bibfnamefont {G.}~\bibnamefont {Moody}}, \bibinfo {author} {\bibfnamefont {H.}~\bibnamefont {Li}}, \bibinfo {author} {\bibfnamefont {A.~D.}\ \bibnamefont {Bristow}}, \ and\ \bibinfo {author} {\bibfnamefont {S.~T.}\ \bibnamefont {Cundiff}},\ }\href {\doibase 10.1364/OE.18.017699} {\bibfield  {journal} {\bibinfo  {journal} {Opt Express}\ }\textbf {\bibinfo {volume} {18}},\ \bibinfo {pages} {17699} (\bibinfo {year} {2010})}\BibitemShut {NoStop}%
\bibitem [{\citenamefont {Martin}\ \emph {et~al.}(2020)\citenamefont {Martin}, \citenamefont {Horng}, \citenamefont {Ruth}, \citenamefont {Paik}, \citenamefont {Wentzel}, \citenamefont {Deng},\ and\ \citenamefont {Cundiff}}]{2020_E.Martin_PRApp_2DCSMoSe2}%
  \BibitemOpen
  \bibfield  {author} {\bibinfo {author} {\bibfnamefont {E.~W.}\ \bibnamefont {Martin}}, \bibinfo {author} {\bibfnamefont {J.}~\bibnamefont {Horng}}, \bibinfo {author} {\bibfnamefont {H.~G.}\ \bibnamefont {Ruth}}, \bibinfo {author} {\bibfnamefont {E.}~\bibnamefont {Paik}}, \bibinfo {author} {\bibfnamefont {M.-H.}\ \bibnamefont {Wentzel}}, \bibinfo {author} {\bibfnamefont {H.}~\bibnamefont {Deng}}, \ and\ \bibinfo {author} {\bibfnamefont {S.~T.}\ \bibnamefont {Cundiff}},\ }\href {\doibase 10.1103/physrevapplied.14.021002} {\bibfield  {journal} {\bibinfo  {journal} {Physical Review Applied}\ }\textbf {\bibinfo {volume} {14}},\ \bibinfo {pages} {021002} (\bibinfo {year} {2020})}\BibitemShut {NoStop}%
\bibitem [{\citenamefont {Efimkin}\ \emph {et~al.}(2021)\citenamefont {Efimkin}, \citenamefont {Laird}, \citenamefont {Levinsen}, \citenamefont {Parish},\ and\ \citenamefont {MacDonald}}]{2021_Dmitry_PRB_theory}%
  \BibitemOpen
  \bibfield  {author} {\bibinfo {author} {\bibfnamefont {D.~K.}\ \bibnamefont {Efimkin}}, \bibinfo {author} {\bibfnamefont {E.~K.}\ \bibnamefont {Laird}}, \bibinfo {author} {\bibfnamefont {J.}~\bibnamefont {Levinsen}}, \bibinfo {author} {\bibfnamefont {M.~M.}\ \bibnamefont {Parish}}, \ and\ \bibinfo {author} {\bibfnamefont {A.~H.}\ \bibnamefont {MacDonald}},\ }\href {\doibase 10.1103/physrevb.103.075417} {\bibfield  {journal} {\bibinfo  {journal} {Physical Review B}\ }\textbf {\bibinfo {volume} {103}},\ \bibinfo {pages} {075417} (\bibinfo {year} {2021})}\BibitemShut {NoStop}%
\bibitem [{\citenamefont {Ngampruetikorn}\ \emph {et~al.}(2012)\citenamefont {Ngampruetikorn}, \citenamefont {Levinsen},\ and\ \citenamefont {Parish}}]{Ngampruetikorn2012}%
  \BibitemOpen
  \bibfield  {author} {\bibinfo {author} {\bibfnamefont {V.}~\bibnamefont {Ngampruetikorn}}, \bibinfo {author} {\bibfnamefont {J.}~\bibnamefont {Levinsen}}, \ and\ \bibinfo {author} {\bibfnamefont {M.~M.}\ \bibnamefont {Parish}},\ }\href {\doibase 10.1209/0295-5075/98/30005} {\bibfield  {journal} {\bibinfo  {journal} {Europhysics Letters}\ }\textbf {\bibinfo {volume} {98}},\ \bibinfo {pages} {30005} (\bibinfo {year} {2012})}\BibitemShut {NoStop}%
\bibitem [{\citenamefont {Hao}\ \emph {et~al.}(2016)\citenamefont {Hao}, \citenamefont {Xu}, \citenamefont {Nagler}, \citenamefont {Singh}, \citenamefont {Tran}, \citenamefont {Dass}, \citenamefont {Schüller}, \citenamefont {Korn}, \citenamefont {Li},\ and\ \citenamefont {Moody}}]{2016_H.Kai_NanoLett_CP-2DCS}%
  \BibitemOpen
  \bibfield  {author} {\bibinfo {author} {\bibfnamefont {K.}~\bibnamefont {Hao}}, \bibinfo {author} {\bibfnamefont {L.}~\bibnamefont {Xu}}, \bibinfo {author} {\bibfnamefont {P.}~\bibnamefont {Nagler}}, \bibinfo {author} {\bibfnamefont {A.}~\bibnamefont {Singh}}, \bibinfo {author} {\bibfnamefont {K.}~\bibnamefont {Tran}}, \bibinfo {author} {\bibfnamefont {C.~K.}\ \bibnamefont {Dass}}, \bibinfo {author} {\bibfnamefont {C.}~\bibnamefont {Schüller}}, \bibinfo {author} {\bibfnamefont {T.}~\bibnamefont {Korn}}, \bibinfo {author} {\bibfnamefont {X.}~\bibnamefont {Li}}, \ and\ \bibinfo {author} {\bibfnamefont {G.}~\bibnamefont {Moody}},\ }\href {\doibase 10.1021/acs.nanolett.6b02041} {\bibfield  {journal} {\bibinfo  {journal} {Nano Letters}\ }\textbf {\bibinfo {volume} {16}},\ \bibinfo {pages} {5109} (\bibinfo {year} {2016})}\BibitemShut {NoStop}%
\bibitem [{\citenamefont {{VanEngen Spivey}}\ \emph {et~al.}(2008)\citenamefont {{VanEngen Spivey}}, \citenamefont {Borca},\ and\ \citenamefont {Cundiff}}]{2008_VanEngen_GaAsCorrelationCoefficient}%
  \BibitemOpen
  \bibfield  {author} {\bibinfo {author} {\bibfnamefont {A.~G.}\ \bibnamefont {{VanEngen Spivey}}}, \bibinfo {author} {\bibfnamefont {C.~N.}\ \bibnamefont {Borca}}, \ and\ \bibinfo {author} {\bibfnamefont {S.~T.}\ \bibnamefont {Cundiff}},\ }\href {\doibase https://doi.org/10.1016/j.ssc.2007.11.006} {\bibfield  {journal} {\bibinfo  {journal} {Solid State Communications}\ }\textbf {\bibinfo {volume} {145}},\ \bibinfo {pages} {303} (\bibinfo {year} {2008})}\BibitemShut {NoStop}%
\bibitem [{\citenamefont {Rodek}\ \emph {et~al.}(2024)\citenamefont {Rodek}, \citenamefont {Oreszczuk}, \citenamefont {Kazimierczuk}, \citenamefont {Howarth}, \citenamefont {Taniguchi}, \citenamefont {Watanabe}, \citenamefont {Potemski},\ and\ \citenamefont {Kossacki}}]{rodek_interactions_2024}%
  \BibitemOpen
  \bibfield  {author} {\bibinfo {author} {\bibfnamefont {A.}~\bibnamefont {Rodek}}, \bibinfo {author} {\bibfnamefont {K.}~\bibnamefont {Oreszczuk}}, \bibinfo {author} {\bibfnamefont {T.}~\bibnamefont {Kazimierczuk}}, \bibinfo {author} {\bibfnamefont {J.}~\bibnamefont {Howarth}}, \bibinfo {author} {\bibfnamefont {T.}~\bibnamefont {Taniguchi}}, \bibinfo {author} {\bibfnamefont {K.}~\bibnamefont {Watanabe}}, \bibinfo {author} {\bibfnamefont {M.}~\bibnamefont {Potemski}}, \ and\ \bibinfo {author} {\bibfnamefont {P.}~\bibnamefont {Kossacki}},\ }\href {\doibase 10.1515/nanoph-2023-0913} {\bibfield  {journal} {\bibinfo  {journal} {Nanophotonics}\ }\textbf {\bibinfo {volume} {13}},\ \bibinfo {pages} {487} (\bibinfo {year} {2024})}\BibitemShut {NoStop}%
\bibitem [{\citenamefont {Zhang}\ \emph {et~al.}(2015)\citenamefont {Zhang}, \citenamefont {You}, \citenamefont {Zhao},\ and\ \citenamefont {Heinz}}]{zhang_experimental_2015}%
  \BibitemOpen
  \bibfield  {author} {\bibinfo {author} {\bibfnamefont {X.-X.}\ \bibnamefont {Zhang}}, \bibinfo {author} {\bibfnamefont {Y.}~\bibnamefont {You}}, \bibinfo {author} {\bibfnamefont {S.~Y.~F.}\ \bibnamefont {Zhao}}, \ and\ \bibinfo {author} {\bibfnamefont {T.~F.}\ \bibnamefont {Heinz}},\ }\href {\doibase 10.1103/PhysRevLett.115.257403} {\bibfield  {journal} {\bibinfo  {journal} {Phys. Rev. Lett.}\ }\textbf {\bibinfo {volume} {115}},\ \bibinfo {pages} {257403} (\bibinfo {year} {2015})}\BibitemShut {NoStop}%
\bibitem [{\citenamefont {Li}\ \emph {et~al.}(2019{\natexlab{a}})\citenamefont {Li}, \citenamefont {Wang}, \citenamefont {Jin}, \citenamefont {Lu}, \citenamefont {Lian}, \citenamefont {Meng}, \citenamefont {Blei}, \citenamefont {Gao}, \citenamefont {Taniguchi}, \citenamefont {Watanabe}, \citenamefont {Ren}, \citenamefont {Cao}, \citenamefont {Tongay}, \citenamefont {Smirnov}, \citenamefont {Zhang},\ and\ \citenamefont {Shi}}]{li_momentum-dark_2019}%
  \BibitemOpen
  \bibfield  {author} {\bibinfo {author} {\bibfnamefont {Z.}~\bibnamefont {Li}}, \bibinfo {author} {\bibfnamefont {T.}~\bibnamefont {Wang}}, \bibinfo {author} {\bibfnamefont {C.}~\bibnamefont {Jin}}, \bibinfo {author} {\bibfnamefont {Z.}~\bibnamefont {Lu}}, \bibinfo {author} {\bibfnamefont {Z.}~\bibnamefont {Lian}}, \bibinfo {author} {\bibfnamefont {Y.}~\bibnamefont {Meng}}, \bibinfo {author} {\bibfnamefont {M.}~\bibnamefont {Blei}}, \bibinfo {author} {\bibfnamefont {M.}~\bibnamefont {Gao}}, \bibinfo {author} {\bibfnamefont {T.}~\bibnamefont {Taniguchi}}, \bibinfo {author} {\bibfnamefont {K.}~\bibnamefont {Watanabe}}, \bibinfo {author} {\bibfnamefont {T.}~\bibnamefont {Ren}}, \bibinfo {author} {\bibfnamefont {T.}~\bibnamefont {Cao}}, \bibinfo {author} {\bibfnamefont {S.}~\bibnamefont {Tongay}}, \bibinfo {author} {\bibfnamefont {D.}~\bibnamefont {Smirnov}}, \bibinfo {author} {\bibfnamefont {L.}~\bibnamefont {Zhang}}, \ and\ \bibinfo {author} {\bibfnamefont {S.-F.}\ \bibnamefont {Shi}},\ }\href {\doibase
  10.1021/acsnano.9b06682} {\bibfield  {journal} {\bibinfo  {journal} {ACS Nano}\ }\textbf {\bibinfo {volume} {13}},\ \bibinfo {pages} {14107} (\bibinfo {year} {2019}{\natexlab{a}})}\BibitemShut {NoStop}%
\bibitem [{\citenamefont {Robert}\ \emph {et~al.}(2017)\citenamefont {Robert}, \citenamefont {Amand}, \citenamefont {Cadiz}, \citenamefont {Lagarde}, \citenamefont {Courtade}, \citenamefont {Manca}, \citenamefont {Taniguchi}, \citenamefont {Watanabe}, \citenamefont {Urbaszek},\ and\ \citenamefont {Marie}}]{robert_fine_2017}%
  \BibitemOpen
  \bibfield  {author} {\bibinfo {author} {\bibfnamefont {C.}~\bibnamefont {Robert}}, \bibinfo {author} {\bibfnamefont {T.}~\bibnamefont {Amand}}, \bibinfo {author} {\bibfnamefont {F.}~\bibnamefont {Cadiz}}, \bibinfo {author} {\bibfnamefont {D.}~\bibnamefont {Lagarde}}, \bibinfo {author} {\bibfnamefont {E.}~\bibnamefont {Courtade}}, \bibinfo {author} {\bibfnamefont {M.}~\bibnamefont {Manca}}, \bibinfo {author} {\bibfnamefont {T.}~\bibnamefont {Taniguchi}}, \bibinfo {author} {\bibfnamefont {K.}~\bibnamefont {Watanabe}}, \bibinfo {author} {\bibfnamefont {B.}~\bibnamefont {Urbaszek}}, \ and\ \bibinfo {author} {\bibfnamefont {X.}~\bibnamefont {Marie}},\ }\href {\doibase 10.1103/PhysRevB.96.155423} {\bibfield  {journal} {\bibinfo  {journal} {Physical Review B}\ }\textbf {\bibinfo {volume} {96}},\ \bibinfo {pages} {155423} (\bibinfo {year} {2017})}\BibitemShut {NoStop}%
\bibitem [{\citenamefont {Molas}\ \emph {et~al.}(2019)\citenamefont {Molas}, \citenamefont {Slobodeniuk}, \citenamefont {Kazimierczuk}, \citenamefont {Nogajewski}, \citenamefont {Bartos}, \citenamefont {Kapuściński}, \citenamefont {Oreszczuk}, \citenamefont {Watanabe}, \citenamefont {Taniguchi}, \citenamefont {Faugeras}, \citenamefont {Kossacki}, \citenamefont {Basko},\ and\ \citenamefont {Potemski}}]{molas_probing_2019}%
  \BibitemOpen
  \bibfield  {author} {\bibinfo {author} {\bibfnamefont {M.}~\bibnamefont {Molas}}, \bibinfo {author} {\bibfnamefont {A.}~\bibnamefont {Slobodeniuk}}, \bibinfo {author} {\bibfnamefont {T.}~\bibnamefont {Kazimierczuk}}, \bibinfo {author} {\bibfnamefont {K.}~\bibnamefont {Nogajewski}}, \bibinfo {author} {\bibfnamefont {M.}~\bibnamefont {Bartos}}, \bibinfo {author} {\bibfnamefont {P.}~\bibnamefont {Kapuściński}}, \bibinfo {author} {\bibfnamefont {K.}~\bibnamefont {Oreszczuk}}, \bibinfo {author} {\bibfnamefont {K.}~\bibnamefont {Watanabe}}, \bibinfo {author} {\bibfnamefont {T.}~\bibnamefont {Taniguchi}}, \bibinfo {author} {\bibfnamefont {C.}~\bibnamefont {Faugeras}}, \bibinfo {author} {\bibfnamefont {P.}~\bibnamefont {Kossacki}}, \bibinfo {author} {\bibfnamefont {D.}~\bibnamefont {Basko}}, \ and\ \bibinfo {author} {\bibfnamefont {M.}~\bibnamefont {Potemski}},\ }\href {\doibase 10.1103/PhysRevLett.123.096803} {\bibfield  {journal} {\bibinfo  {journal} {Physical Review Letters}\ }\textbf {\bibinfo {volume}
  {123}},\ \bibinfo {pages} {096803} (\bibinfo {year} {2019})}\BibitemShut {NoStop}%
\bibitem [{\citenamefont {Zhou}\ \emph {et~al.}(2017)\citenamefont {Zhou}, \citenamefont {Scuri}, \citenamefont {Wild}, \citenamefont {High}, \citenamefont {Dibos}, \citenamefont {Jauregui}, \citenamefont {Shu}, \citenamefont {De~Greve}, \citenamefont {Pistunova}, \citenamefont {Joe}, \citenamefont {Taniguchi}, \citenamefont {Watanabe}, \citenamefont {Kim}, \citenamefont {Lukin},\ and\ \citenamefont {Park}}]{zhou_probing_2017}%
  \BibitemOpen
  \bibfield  {author} {\bibinfo {author} {\bibfnamefont {Y.}~\bibnamefont {Zhou}}, \bibinfo {author} {\bibfnamefont {G.}~\bibnamefont {Scuri}}, \bibinfo {author} {\bibfnamefont {D.~S.}\ \bibnamefont {Wild}}, \bibinfo {author} {\bibfnamefont {A.~A.}\ \bibnamefont {High}}, \bibinfo {author} {\bibfnamefont {A.}~\bibnamefont {Dibos}}, \bibinfo {author} {\bibfnamefont {L.~A.}\ \bibnamefont {Jauregui}}, \bibinfo {author} {\bibfnamefont {C.}~\bibnamefont {Shu}}, \bibinfo {author} {\bibfnamefont {K.}~\bibnamefont {De~Greve}}, \bibinfo {author} {\bibfnamefont {K.}~\bibnamefont {Pistunova}}, \bibinfo {author} {\bibfnamefont {A.~Y.}\ \bibnamefont {Joe}}, \bibinfo {author} {\bibfnamefont {T.}~\bibnamefont {Taniguchi}}, \bibinfo {author} {\bibfnamefont {K.}~\bibnamefont {Watanabe}}, \bibinfo {author} {\bibfnamefont {P.}~\bibnamefont {Kim}}, \bibinfo {author} {\bibfnamefont {M.~D.}\ \bibnamefont {Lukin}}, \ and\ \bibinfo {author} {\bibfnamefont {H.}~\bibnamefont {Park}},\ }\href {\doibase 10.1038/nnano.2017.106} {\bibfield
   {journal} {\bibinfo  {journal} {Nature Nanotechnology}\ }\textbf {\bibinfo {volume} {12}},\ \bibinfo {pages} {856} (\bibinfo {year} {2017})}\BibitemShut {NoStop}%
\bibitem [{\citenamefont {Li}\ \emph {et~al.}(2019{\natexlab{b}})\citenamefont {Li}, \citenamefont {Wang}, \citenamefont {Lu}, \citenamefont {Khatoniar}, \citenamefont {Lian}, \citenamefont {Meng}, \citenamefont {Blei}, \citenamefont {Taniguchi}, \citenamefont {Watanabe}, \citenamefont {McGill}, \citenamefont {Tongay}, \citenamefont {Menon}, \citenamefont {Smirnov},\ and\ \citenamefont {Shi}}]{li_direct_2019}%
  \BibitemOpen
  \bibfield  {author} {\bibinfo {author} {\bibfnamefont {Z.}~\bibnamefont {Li}}, \bibinfo {author} {\bibfnamefont {T.}~\bibnamefont {Wang}}, \bibinfo {author} {\bibfnamefont {Z.}~\bibnamefont {Lu}}, \bibinfo {author} {\bibfnamefont {M.}~\bibnamefont {Khatoniar}}, \bibinfo {author} {\bibfnamefont {Z.}~\bibnamefont {Lian}}, \bibinfo {author} {\bibfnamefont {Y.}~\bibnamefont {Meng}}, \bibinfo {author} {\bibfnamefont {M.}~\bibnamefont {Blei}}, \bibinfo {author} {\bibfnamefont {T.}~\bibnamefont {Taniguchi}}, \bibinfo {author} {\bibfnamefont {K.}~\bibnamefont {Watanabe}}, \bibinfo {author} {\bibfnamefont {S.~A.}\ \bibnamefont {McGill}}, \bibinfo {author} {\bibfnamefont {S.}~\bibnamefont {Tongay}}, \bibinfo {author} {\bibfnamefont {V.~M.}\ \bibnamefont {Menon}}, \bibinfo {author} {\bibfnamefont {D.}~\bibnamefont {Smirnov}}, \ and\ \bibinfo {author} {\bibfnamefont {S.-F.}\ \bibnamefont {Shi}},\ }\href {\doibase 10.1021/acs.nanolett.9b02132} {\bibfield  {journal} {\bibinfo  {journal} {Nano Letters}\ }\textbf {\bibinfo
  {volume} {19}},\ \bibinfo {pages} {6886} (\bibinfo {year} {2019}{\natexlab{b}})}\BibitemShut {NoStop}%
\end{thebibliography}%

\clearpage


\end{document}